\documentclass[aps,prc,groupedaddress,twocolumn,showpacs,amsmath,amssymb,floatfix]{revtex4}
\usepackage{graphics}% Include figure files
\usepackage{dcolumn}% Align table columns on decimal point 
\usepackage{longtable}

\begin{document}

\title{Optical potential for incident and emitted low-energy $\alpha$ particles. III. Non-statistical processes induced by neutrons on Zr, Nb, and Mo nuclei}
\author{M.~Avrigeanu} \email{marilena.avrigeanu@nipne.ro}
\author{V.~Avrigeanu} \email{vlad.avrigeanu@nipne.ro}
\affiliation{Horia Hulubei National Institute for Physics and Nuclear Engineering, 077125 Bucharest-Magurele, Romania}
%\date{\today}

\begin{abstract}
%\begin{description} \normalfont\large\Large
\noindent
{\bf Background:} The reliability of a previous $\alpha$-particle optical-model potential (OMP) on nuclei with mass number 45$\leq$$A$$\leq$209 was proved for emitted $\alpha$ particles as well, for proton--induced reactions on Zn isotopes [Phys. Rev. C {\bf 91}, 064611 (2015), Paper I]. However, the same was not the case of neutrons on Zr stable isotopes [Phys. Rev. C {\bf 96}, 044610 (2017), Paper II]. 
\smallskip

\noindent
{\bf Purpose:} 
A recent assessment of this potential also for nucleon--induced $\alpha$-emission on $A$$\sim$60 nuclei, including pickup direct reaction and eventual Giant Quadrupole Resonance (GQR) $\alpha$-emission, is completed for neutrons incident on Zr, Nb, and Mo stable isotopes. 

\smallskip

\noindent
{\bf Methods:} Consistent sets of input parameters, determined through analysis of independent data, is involved %at variance with the use of either empirical rescaling factors of the $\gamma$ and neutron widths or combinations of all computer--code options for main input parameters.
while no further empirical rescaling factors of the $\gamma$ and nucleon widths have been involved. 

\smallskip

\noindent
{\bf Results:} 
A suitable account of all competitive reaction channels is confirmed by careful uncertainty analysis, to avoid parameter ambiguities and/or error compensation. 
Additional validation of this potential is also supported by recently measured  $(\alpha,\gamma)$ and $(\alpha,n)$ cross--sections of Zr and Mo nuclei. 
%Nevertheless, there is an obvious correlation between the accuracy of the independent data, the input parameters determined by their fit, and the final uncertainties of the calculated reaction cross sections. 
 
\smallskip

\noindent
{\bf Conclusions}: 
An increase of the $\alpha$-emission beyond the statistical predictions, through consideration of additional reaction channels of the pickup direct interaction and {\it like}--GQR decay, makes possible the description of both absorption and emission of $\alpha$ particles by the same optical potential. 
%\end{description}
\end{abstract}


\maketitle
 
\section{Introduction}
\label{intro}
The reliability of a previous $\alpha$-particle optical-model potential (OMP) for target nuclei with mass number 45$\leq$$A$$\leq$209 \cite{va14} was also proved for $\alpha$-emission in proton--induced reactions on Zn isotopes \cite{va15} (Paper I). 
However, its use led to underestimated predictions of the statistical Hauser-Feshbach (HF) \cite{wh52} and pre-equilibrium emission (PE) \cite{eg92} models for neutrons incident on Zr stable isotopes \cite{va17} (Paper II). 
On the other hand, it has recently been shown that this potential can also describe $\alpha$-emission from excited nuclei in nucleon-induced reactions within the $A$$\sim$60 range \cite{va21,va22}. 
This was possible through additional consideration of the pickup direct reaction (DR) and eventual Giant Quadrupole Resonance (GQR) $\alpha$-emission. 
The so-called $\alpha$-potential mystery \cite{tr13} related to the account of both $\alpha$-emission and absorption by an OMP, of equal interest for astrophysics and fusion technology, thus received an alternate solution. 
Nevertheless, its support by analysis of more data is imperative, while the case of neutrons incident on $A$$\sim$90 nuclei \cite{va17} is a distinct requisite in this respect. 

Thus, the same analysis for the stable isotopes of   neighboring elements Zr, Nb, and Mo becomes challenging, with Zr nuclei within both incident and emergent reaction channels. 
The $\alpha$-emission in neutron--induced reactions on stable Mo isotopes was targeted in an earlier systematic investigation up to 20 MeV \cite{pr05}. 
However, an $\alpha$-particle OMP \cite{va94} describing $A$$\sim$60 compound--nuclei (CN) $\alpha$-particle decay led to a significant underestimation at incident energies around $\leq$10 MeV for $^{92,98}$Mo. 
This OMP makes predictions that differ significantly from potentials for incident $\alpha$ particles \cite{lmf66} including a double--folding (DFM) microscopic real potential \cite{va14,ma03,ma06,ma09,ma10}. 
The discrepancy in results corresponding to OMPs \cite{va94,ma03} led even to the  assumption of a nuclear--density distribution dependent on nuclear temperature \cite{ma06}. 
%However, still open questions of the $\alpha$-emission insight at that time pointed out as a first aim the validation of a certain OMP for incident $\alpha$-particles.

The removal of compensation effects of less accurate model parameters is essential to establishing an appropriate $\alpha$-particle  OMP \cite{va14,ma03,ma06,ma09,ma10} or to prove it \cite{va16,va19}. 
This aim has been achieved using "consistent sets of input parameters, determined through analysis of independent data available in this mass region" \cite{eda80}. 
Moreover, a proper account of all competitive reaction channels, beyond the $\alpha$-emission of interest, should also be aimed at the validation of consistent parameter sets. 
So, there have been avoided empirical rescaling factors of the $\gamma$ and/or nucleon widths which however are mandatory within large-scale nuclear--data evaluations. 

This work is an extension of Refs. \cite{va17,va21,va22} so that only additional HF+PE model parameters for Nb and Mo isotopes are given in Sec.~\ref{SMcalc} along with items of the DR analysis by the distorted-wave Born approximation (DWBA) method and code FRESCO \cite{FRESCO}. 
A comparison of HF+PE results and (i) recent $(\alpha,\gamma)$ and $(\alpha,n)$ cross--sections of Zr and Mo nuclei, for $\alpha$-particle OPM additional proof, (ii) available data of neutron--induced activation by nucleon--emission for $^{93}$Nb and $^{92}$Mo, as for Zr nuclei \cite{va17}, and (iii) $\alpha$-emission for all Zr, Nb, and Mo isotopes including eventual excited--nucleus {\it like}--GQR decay, are discussed in Sec.~\ref{Res}. 
Conclusions are finally given in Sec.~\ref{Conc}. 

\begingroup
\squeezetable
\begin{table*} % add [H] placement to break table across pages
\caption{\label{densp} Low-lying levels number $N_d$ up to excitation energy $E^*_d$ \protect\cite{ensdf} used in HF calculations of reaction cross--sections, low-lying levels and $s$-wave nucleon-resonance spacings$^a$ $D_0^{\it exp}$ (with uncertainties given in units of the last digit in parentheses) in the energy range $\Delta$$E$ above the separation energy $S$, for the target-nucleus ground state (g.s.) spin $I_0$, fitted to obtain the BSFG level-density parameter {\it a} and g.s. shift $\Delta$ (for a spin cutoff factor calculated with a variable moment of inertia \cite{va02} between half and 75\% of the rigid-body value, from g.s. to $S$, and reduced radius $r_0$=1.25 fm).} 
\begin{ruledtabular}
\begin{tabular}{cccccccccc} 
Nucleus   &$N_d$&$E^*_d$& \multicolumn{5}{c}
                  {Fitted level and resonance data}& $a$ & $\Delta$\hspace*{3mm}\\
\cline{4-8}
           &  &     &$N_d$&$E^*_d$&$S+\frac{\Delta E}{2}$&
                                     $I_0$&$D_0^{\it exp}$ \\ 
           &  &(MeV)&   & (MeV)& (MeV)&  &(keV)&(MeV$^{-1}$) & (MeV) \\ 
\noalign{\smallskip}\hline\noalign{\smallskip}
$^{87}$Sr&29&2.708&53(2)&3.166& 8.442 & 0 &  2.6(8)  & 9.16(53/-36)(49/-33)& 0.09(19/-14)(15/-10) \\   % 2015
$^{88}$Sr&47&4.801&47   &4.801&11.113 &9/2& 0.29(8)  & 8.70& 1.63 \\   % 2014
$^{89}$Sr&22&3.249&22(2)&3.249& 6.430 & 0 & 23.7(29) & 9.58(24/-19)(15/-0)& 0.87(7/-5)(-1/4) \\   % 2013 
$^{90}$Sr&15&3.039&17   &3.146&       &   &          & 9.60& 0.95 \\
$^{91}$Sr&15&2.237&15(3)&2.237&       &   &          & 9.7(3)& 0.13(7)(5) \\   % 2013 
$^{92}$Sr&15&2.925&15   &2.925&       &   &          &10.00& 0.89 \\
$^{93}$Sr&20&2.169&20(2)&2.169&       &   &          &10.6(3)& 0.09(7)(1) \\   % 2011   
%$^{94}$Sr&23&3.155&23&3.155&       &   &          &11.00& 1.08 \\   % 2011   

%$^{86}$ Y&21&1.277&21&1.277&       &   &          & 9.40&-1.12 \\ 
%$^{87}$ Y&24&1.849&64&2.502&       &   &          & 9.50&-0.57 \\ 
%$^{88}$ Y&24&1.477&17&1.262&       &   &          & 9.40&-1.12 \\ 
$^{89}$ Y&26&3.630&26   &3.630&11.478 & 4 &0.106(35)$^b$&8.90&0.94\\ 
$^{90}$ Y&30&2.366&29(2)&2.327& 6.857 &1/2&  3.7(4)  & 9.23(15)(10)&-0.32(5)(2) \\ 
%$^{91}$ Y&11&1.580&10&1.547&       &   &          & 9.30&-0.40 \\ 
%$^{92}$ Y& 4&0.431& 4&0.431&       &   &          &10.40&-1.00 \\ 
%$^{93}$ Y&22&2.200&21&2.129&       &   &          &10.10&-0.08 \\ 
%$^{94}$ Y& 4&0.724&[4&0.431]$^b$&  &   &          &11.40&-0.80 \\ 
%$^{95}$ Y&10&2.047&10&2.047&       &   &          &11.40& 0.50 \\ 
%$^{96}$ Y& 3&0.652&[4&0.431]$^b0$&  &   &          &12.00&-0.70 \\ 

%$^{87}$Zr&24&1.949&24&1.949&       &   &          & 9.15&-0.58 \\ 
$^{88}$Zr&22&3.094&22&3.094&       &   &          & 8.90& 0.56 \\   % 2014 
$^{89}$Zr&23&2.300&23(3)&2.300&       &   &          & 9.2(4)&-0.19(13)(4) \\   % 2012 
$^{90}$Zr&46&4.701&46&4.701&       &   &          & 9.00& 1.65 \\   % 2020
$^{91}$Zr&37&3.053&37&3.053& 7.260 & 0 & 6.0(14)  & 9.77& 0.40 \\   % 2013  
$^{92}$Zr&42&3.500&54(2)&3.725& 8.647 &5/2& 0.55(10) & 9.67(27)(25)& 0.79(9)(6) \\   % 2012  
$^{93}$Zr&29&2.391&29&2.391& 6.785 & 0 & 3.5(8)   &10.66& 0.12 \\   % 2011  
$^{94}$Zr&23&3.059&23&3.059& 8.220 &5/2& 0.302(75)&11.00& 1.00 \\   % 2006  
$^{95}$Zr&14&2.022&14(2)&2.022& 6.507 & 0 & 4.0(8)   &11.04(33)(19)& 0.23(7)(5) \\   % 2010 
$^{96}$Zr&38&3.630&38&3.630&       &   &          &11.20& 1.32 \\   % 2008  
$^{97}$Zr& 9&2.058& 9(1)&2.058& 5.629 & 0 & 13(3)    &11.21(42)(30)& 0.51(7)(2) \\   % 2010 

%$^{89}$Nb&20&2.221&20&2.221&       &   &               & 9.40&-0.12 \\   % 2012 
%$^{90}$Nb&26&1.433&26&1.433&       &   &               & 9.20&-1.17 \\   % 2020 
$^{91}$Nb&29&2.660&29&2.660&       &   &          & 9.30& 0.04 \\   % 2013   
$^{92}$Nb&41&1.851&41&1.851&       &   &          & 9.60&-0.92 \\   % 2012  
$^{93}$Nb&35&1.784&35&1.784&       &   &          & 9.90&-0.80 \\   % 2011 
$^{94}$Nb&48&1.281&48&1.281& 7.232 &9/2&0.094(10) &10.75&-1.29 \\   % +2020
$^{95}$Nb&28&1.660&28&1.660&       &   &          &11.30&-0.45 \\   % 2010
$^{97}$Nb&19&2.113&19&2.113&       &   &          &11.60& 0.26 \\   % 2010
$^{98}$Nb& 3&0.226& 3&0.226&       &   &          &13.50&-0.52 \\   % 2020
$^{99}$Nb&15&1.305&15&1.305&       &   &          &12.40&-0.31 \\   % 2010 
$^{100}$Nb&19&0.772&19&0.772&      &   &          &13.50&-0.48 \\   % 2021 

$^{90}$Mo&22&3.185&22&3.185&       &   &          & 9.00& 0.67 \\   % 2020 
$^{91}$Mo&24&2.345&29&2.716&       &   &          & 9.30& 0.09 \\   % 2013 
$^{92}$Mo&37&4.187&37&4.187&       &   &          & 9.20& 1.36 \\   % 2012  
$^{93}$Mo&60&2.974&77&3.161& 8.092 & 0 &2.7(5),2.17(25)$^c$&9.6&-0.02\\ % 2011 
$^{94}$Mo&60&3.456&61&3.462& 9.678 &5/2&0.081(24)$^d$&10.74& 0.78 \\% 2021  
$^{95}$Mo&25&1.692&25&1.692& 7.377 & 0 & 1.32(18) &10.45&-0.56 \\   % 2010 
%$^{96}$Mo&38&2.875&38&2.875& 9.154 &5/2&0.0661(30)$^c$&11.35&0.61\\ % 2008 
$^{96}$Mo&38&2.875&38&2.875& 9.154 &5/2&0.0662(30)$^e$&11.35&0.61\\ % 2008 
$^{97}$Mo&33&1.341&33&1.341& 6.831 & 0 &1.05(20).0.661(198)$^d$&11.16&-0.91\\% 2010
$^{98}$Mo&38&2.701&38&2.701& 8.664 &5/2&0.06(1),0.047(6)$^d$&12.20(40)(36)&0.60(8)(4)\\% 2020 
$^{99}$Mo&30&1.198&49&1.341& 5.941 & 0 & 1.0(2)   &12.34(37)(32)&-0.75(8)(4)\\% 2017
$^{100}$Mo&32&2.464&31&2.432&      &   &          &12.00& 0.39\\% 2008 
$^{101}$Mo&23&0.626&23&0.626&5.411 & 0 & 0.62(10) &13.14&-1.19\\% 2006
\end{tabular}	 
\end{ruledtabular}
\begin{flushleft}
$^a$RIPL-3 \cite{ripl3} if not otherwise mentioned 
%$^a0$Reference \cite{mg14}\\ 
%$^b0$Levels of $^{92}$Y nucleus\\
\hspace*{0.35in} $^b$Reference \cite{mg14}
\hspace*{0.35in} $^c$RIPL-1 Beijing file \cite{ripl1b} 
\hspace*{0.35in} $^d$Reference \cite{hu13}
\hspace*{0.35in} $^e$Reference \cite{pek22}\\ 
\end{flushleft}
\end{table*}
\endgroup

\section{Models and parameters}
\label{SMcalc}
\subsection{Compound and pre-equilibrium emission}

The same models, codes \cite{TALYS,ma95,pdk84}, and local approaches \cite{va17,va21} concern the HF+PE and collective inelastic--scattering cross--section assessment. 
Thus, consistent (i) back-shifted Fermi gas (BSFG) \cite{hv88} nuclear level density (NLD) parameters, (ii) nucleon and (iii) $\gamma$-ray transmission coefficients have also been involved in this work, with the related parameters being established or validated using distinct data. 
The same OMP and level density parameters have been used in the framework of the HF, PE, and DR models, too. 

The excitation functions calculated in this work are also compared with the results of the worldwide used code TALYS-1.96 \cite{TALYS} and its default options which include the $\alpha$-particle OMP \cite{va14}. 
%This inclusion followed better results obtained within large-scale nuclear-data evaluation, in comparison to the other related options of this code. 
%Thus, the comparison of the $(n,x\alpha)$ cross--sections corresponding to the same $\alpha$-particle OMP in the current work, along with all default options of TALYS-1.96 supports additionally the consistency of the local parameter set. 
%
Furthermore, a correlation with the TALYS–-based evaluated--data library TENDL-2021 \cite{TENDL} shows the evaluation progress vs default options usage. 

%On the other hand a parallel can thus be drawn between the use of consistent parameters and various rescaling factors which are usual in current evaluations. 

\begin{figure*} %[b]
\resizebox{1.95\columnwidth}{!}{\includegraphics{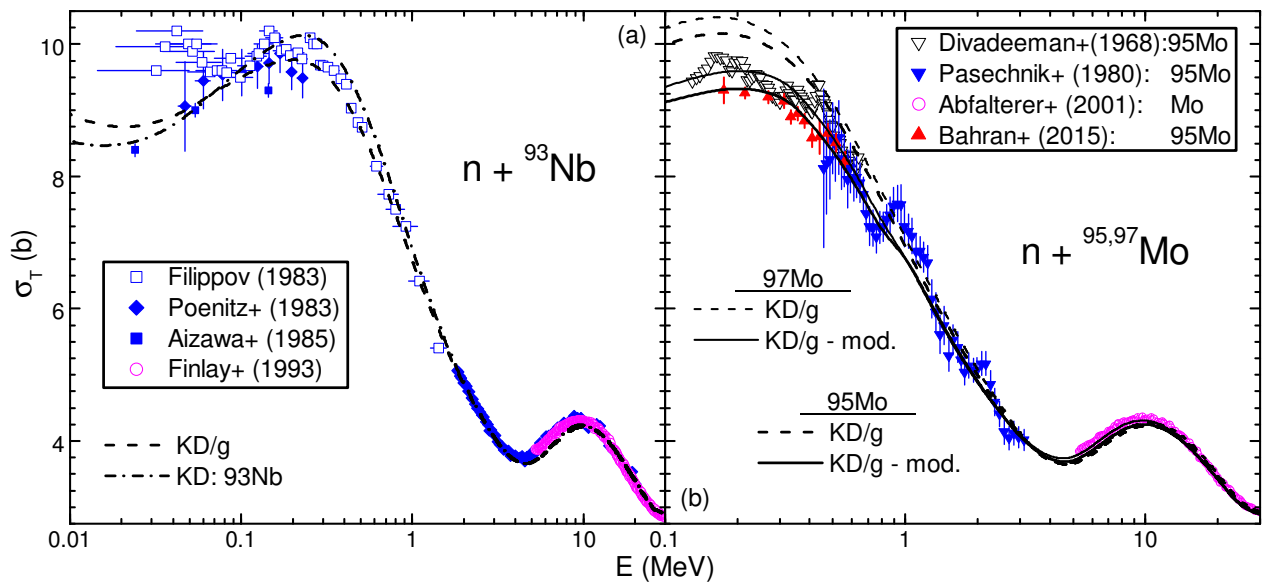}}
\caption{\label{Fig:Nb93Mo950nT} Comparison of measured \cite{exfor} and calculated neutron total cross--sections of (a) $^{93}$Nb, using the global (dashed curve) and local (dash-dotted curve) OMP parameters of Koning and Delaroche \cite{KD03}, and (b) $^{95}$Mo (thick, lower curves) and $^{97}$Mo (thin, upper curves) with global (dashed curves) as well as the corresponding modified values of $v_1$ parameters given in Table~\ref{tab:nres} (solid curves).}
\end{figure*}

{\it The low-lying levels and NLD parameters} of the BSFG model are given in Table~\ref{densp} for all nuclei within this work. 
The limits of the fitted $a$ and $\Delta$ parameters were obtained also by fit of the error-bars of $s$-wave nucleon--resonance spacings $D_0^{\it exp}$. 
For nuclei without resonance data, the smooth-curve method \cite{chj77} was used, in which an average of fitted $a$-values for nearby nuclei is adopted while the $\Delta$ value is obtained by fit of the low-lying discrete levels. 
These NLD parameter limits have then been used to illustrate the NLD effects on HF calculated cross--sections (Sec.~\ref{Res}). 
%The larger uncertainties of averaged $a$-values, due to the spread of the fitted $a$ parameters, lead to increased uncertainty bands of HF cross--sections. 
The additional uncertainty of fitted $N_d$ led to enlarged NLD parameter uncertainties (second pair of brackets in Table~\ref{densp}) despite better data becoming available in the meantime (e.g. for A=90, 98--100 nuclei \cite{ensdf}).

%\subsubsection{Neutron optical model potentials} \label{OMPn}

\begin{figure} %[b]
\resizebox{0.99\columnwidth}{!}{\includegraphics{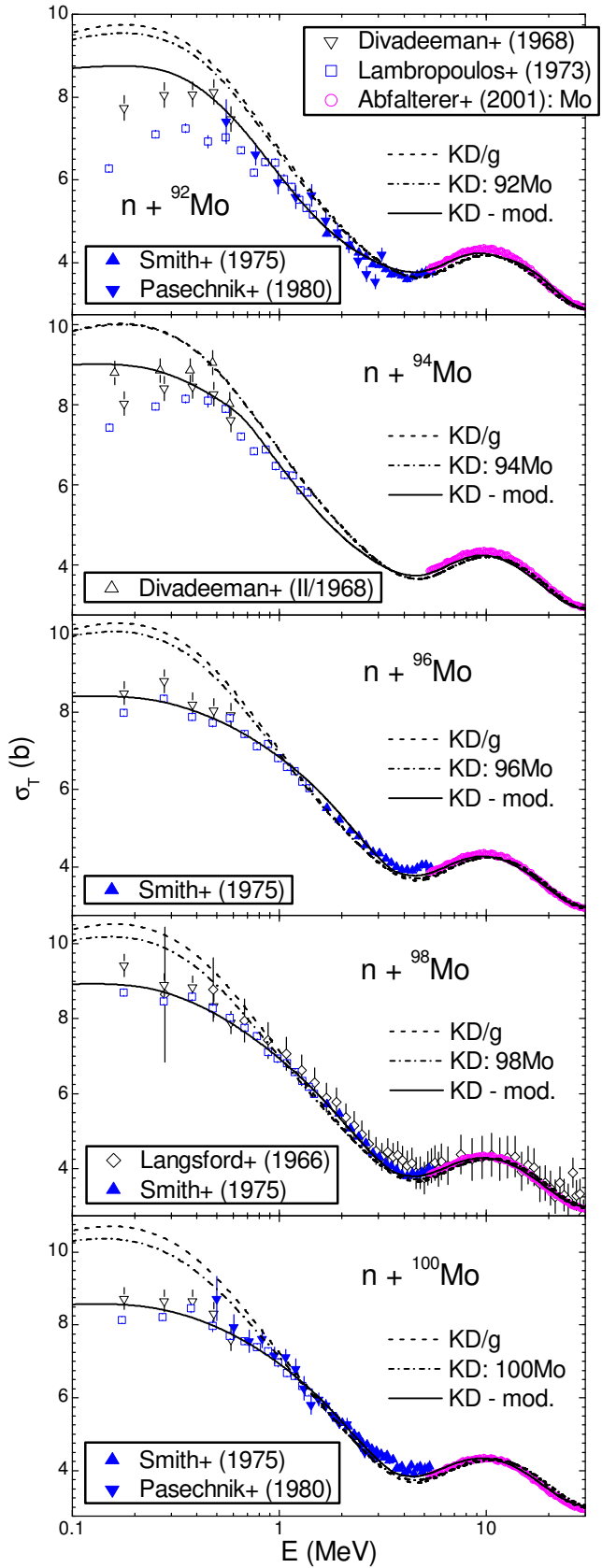}}
\caption{\label{Fig:Mo9246810nT} As Fig.~\ref{Fig:Nb93Mo950nT} but for  $^{92,94,96,98,100}$Mo and modified (Table~\ref{tab:nres}) (solid curves) local real--potential depth parameter $v_1$ \cite{KD03}.} 
\end{figure}

{\it The neutron OMP} of Koning and Delaroche \cite{KD03} has additionally been reviewed at energies up to $\sim$30 MeV through the SPRT method \cite{jpd76}, i.e. by analysis of the $s$- and $p$-wave neutron strength functions $S_0$ and $S_1$, respectively, the potential scattering radius $R'$ \cite{ripl3} (Table~\ref{tab:nres}), and the energy dependence of the neutron total cross--section $\sigma_T (E)$ \cite{exfor}. 
Thus, we found that the global parameter set \cite{KD03} provides a better agreement between the measured and calculated $\sigma_T (E)$ of $^{93}$Nb within this energy range, in comparison to the related local parameter set [Fig.~\ref{Fig:Nb93Mo950nT}(a)]. 

The small adjustments shown in Table~\ref{tab:nres} for the real potential depth parameter $v_1$ \cite{KD03} either global, for the odd $^{95,97}$Mo [Fig.~\ref{Fig:Nb93Mo950nT}(b)], or local for the even Mo isotopes (Fig.~\ref{Fig:Mo9246810nT}), have the same results. 
One may note the neutron energies for this comparison, which are quite different from the usual value of 10 keV. 
These energies are taken from the earlier RIPL-1 compilation \cite{ripl1b} except the case of more recent measurements \cite{pek22,gl10,rb15,ker18}. 
Nevertheless, these changes have no importance for the $\sigma_T (E)$ values around 1 MeV, which are of obvious importance for the competition between the neutron and charged--particle decay of excited CN, in heavier Mo nuclei. 
It is also within $\sim$4\% for the near-spherical $^{92}$Mo nucleus as well as for $^{94}$Mo. 

Because the improvement of the calculated $\sigma_T (E)$ at higher energies is yet within the measured error bars, the present analysis for Nb and Mo isotopes has also supported the previous use \cite{va17} of the OMP \cite{KD03} for Zr isotopes. 

The same OMP parameters have been involved within DWBA assessment of the collective inelastic scattering, using the deformation parameters \cite{ck00,sr01,tk02} of the first 2$^+$ and 3$^-$ collective states. 
Typical direct inelastic--scattering cross--sections increase to, e.g., $\sim$7\% of $\sigma_R$ for neutrons on $^{93}$Nb and $^{98}$Mo at incident energies around 4 MeV and then decrease to slightly less than 5\% at energies above 20 MeV. 
They were then involved in a subsequent decrease of $\sigma_R$ that has been taken into account within the PE+HF analysis. 

%\enlargethispage{}
\begin{table} %[h]
%\squeezetable
\caption{\label{tab:nres}Comparison of experimental \cite{ripl3,pek22,gl10,rb15,ker18} and calculated $s$- and $p$-wave neutron strength functions $S_0$ and $S_1$, respectively, and the potential scattering radius $R'$ \cite{ripl3} of $^{93}$Nb and $^{92,94-98,100}$Mo isotopes, %up to the shown neutron energies, %of $\sim$0.4, $\sim$0.2, 0.034, $\sim$0.28, and 0.3 MeV \cite{ripl1b}, respectively, 
and the changes of either local or global real--potential depth $v_1$ of the even and odd Mo isotopes \cite{KD03}, respectively (with the use of \cite{KD03} notations, the energies are in MeV and geometry parameters in fm) which provide the best results within the SPRT method.}
\begin{ruledtabular}
\begin{tabular}{clccc}%|ccc}
%\hline\noalign{\smallskip} 
Nucleus &\hspace*{0.00in} Reference Exp./OMP & 10$^4$S$_0$ & 10$^4$S$_1$ & R' \\
\hline
$^{93}$Nb &\cite{ripl3} (E$<$0.004 \cite{ripl1b}) & 0.45(7) & 5.8(8)& 7.0(2) \\ 
          &\cite{KD03}, global        	        & 0.55    & 6.48   & 6.4 \\ 
          &\cite{KD03}, local         	        & 0.44    & 6.28   & 6.5 \\ 
\hline
$^{92}$Mo &\cite{ripl3} (E$<$0.025 \cite{ripl1b}) & 0.56(7) & 3.6(6) & 6.4(8) \\ 
          &\cite{KD03}, global        	        & 0.55    & 7.0    & 6.3 \\ 
          &\cite{KD03}, local (v1=55.0)	        & 0.61    & 6.4    & 6.4 \\ 
          &\cite{KD03}+ v1=53.5, E$<$11.5  		& 0.66    & 4.5    & 6.6 \\ 
					&\hspace*{0.57in}57.0\\
\hline
$^{94}$Mo &\cite{ripl3} (E$<$0.01 \cite{ripl1b})  & 0.44(8) & 7.2(27)& 6.5(13) \\ 
          &\cite{KD03}, global        	        & 0.54    & 7.4    & 6.4 \\ 
          &\cite{KD03}, local (v1=54.2)	        & 0.56    & 7.3    & 6.3 \\ 
          &\cite{KD03}+ v1=57.5, E$<$0.65    	& 0.59    & 8.2    & 5.8 \\ 
					&\hspace*{0.57in}53.0\\
\hline
$^{95}$Mo &\cite{ripl3} (E$<$0.0025\cite{ripl1b}) & 0.47(17)& 6.9(18) \\ 
          &\cite{gl10},  E$<$0.001                & 0.436(9) \\ 
          &\cite{rb15},  E$<$0.36                 & 0.4(1)  & 3.8(1) & 6.97(4) \\ 
          &\cite{pek22},  E$<$0.005               & 0.47(17)& 3.09(35) \\ 
          &\cite{KD03}, global (v1=54.6)	      & 0.50    & 5.1    & 6.4 \\ 
          &\cite{KD03}+ v1=53.0, E$<$1     		& 0.52    & 4.5    & 6.2 \\ 
					&\hspace*{0.57in}54.0\\
\hline
$^{96}$Mo &\cite{ripl3} (E$<$0.01 \cite{ripl1b})  & 0.62(12)& 7.1(16)& 6.6(13) \\
          &\cite{KD03}, global        	        & 0.53    & 8.0    & 6.3 \\  
          &\cite{KD03}, local (v1=55.0)	        & 0.56    & 7.7    & 6.3 \\ 
          &\cite{KD03}+ v1=57.2, E$<$3.5   		& 0.62    & 7.2    & 5.7 \\ 
					&\hspace*{0.57in}52.5\\
\hline
$^{97}$Mo &\cite{ripl3} (E$<$0.002 \cite{ripl1b}) & 0.37(7) & 7.85(153) \\ 
          &\cite{gl10}, E$<$0.001                 & 0.438(5) \\ 
          &\cite{KD03}, global (v1=54.2)	      & 0.53    & 8.2    & 6.4 \\ 
          &\cite{KD03}+ v1=52.5, E$<$0.7       & 0.55    & 5.5    & 6.6 \\ 
					&\hspace*{0.57in}53.0, E$>$1.0 \\
\hline
$^{98}$Mo &\cite{ripl3} (E$<$0.016 \cite{ripl1b}) & 0.48(9) & 6.3(21) & 7.0(14) \\ 
          &\cite{ker18}, E$<$0.005                & 0.7(4)  & 6.2(11) &  \\ 
          &\cite{ker18}, E$<$0.028                & 0.4(2)  & 7.4(10) &  \\ 
          &\cite{KD03}, global        	        & 0.51(1) & 8.5     & 6.3 \\ 
          &\cite{KD03}, local (v1=54.0)	        & 0.54    & 8.1     & 6.3 \\ 
          &\cite{KD03}+ v1=56.0, E$<$3.5   	  & 0.59(1) & 7.7(2)  & 5.8(1) \\ 
					&\hspace*{0.57in}52.0\\
\hline
$^{100}$Mo&\cite{ripl3} (E$<$0.013 \cite{ripl1b}) & 0.7(1)  & 5.0(8) & 6.5(13) \\ 
          &\cite{ker18}, E$<$0.005                & 0.4(2)  & \\ 
          &\cite{ker18}, E$<$0.021                & 0.9(2)  & 4.2 (6) \\ 
          &\cite{KD03}, global        	        & 0.50    & 9.0     & 6.3 \\ 
          &\cite{KD03}, local (v1=52.0)	        & 0.52(1) & 8.8 (1) & 6.2 (1) \\ 
          &\cite{KD03}+ v1=55.5, E$<$3.5   		& 0.58(1) & 7.2(2)  & 5.8 \\ 
					&\hspace*{0.57in}51.0
%\hline
%\noalign{\smallskip}\hline 
\end{tabular}
\end{ruledtabular}
%}
\end{table}

\begin{figure*} %[h]
\resizebox{1.50\columnwidth}{!}{\includegraphics{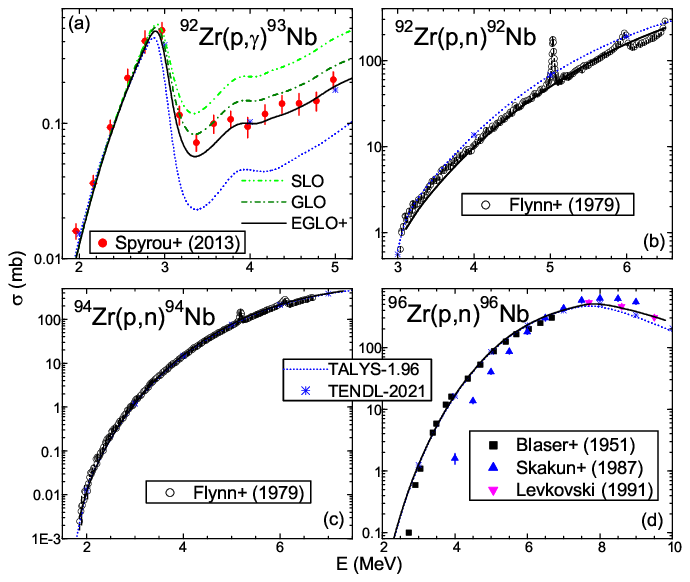}}
\caption{\label{Fig:Zr9246pgn} Comparison of cross--sections measured \cite{exfor}, evaluated \cite{TENDL} ($\ast$), code TALYS-1.96 \cite{TALYS} default--option results (short--dotted curves), and calculated in this work using proton global OMP parameters \cite{KD03} and $E$1-radiation EGLO strength functions(solid curves) for (a) $(p,\gamma)$ reaction on $^{92}$Zr, including results for alternate use of $E$1-radiation GLO (dash-dotted) and SLO (dash-dot-dotted) models, and (b--d) $(p,n)$ reaction on $^{92,94,96}$Zr, respectively.}
\end{figure*}

\begin{figure} %[h]
\resizebox{1.0\columnwidth}{!}{\includegraphics{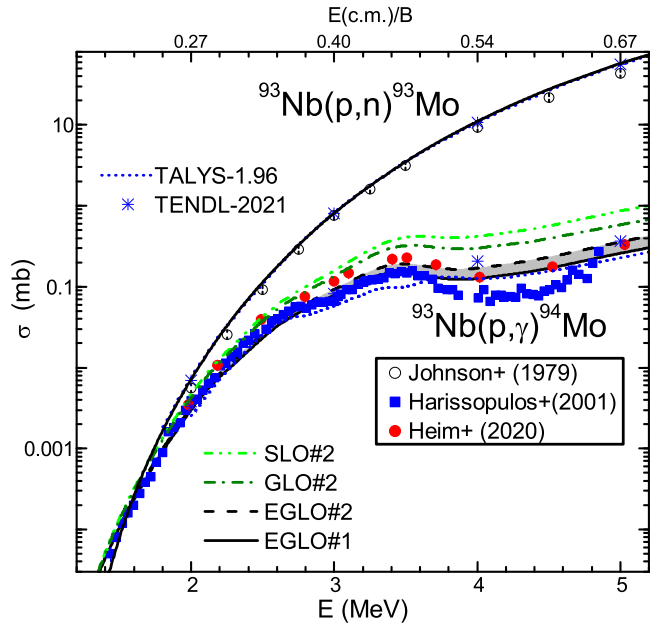}}
\caption{\label{Fig:Nb93pgn} As Fig.~\ref{Fig:Zr9246pgn} but for $(p,n)$ reaction (upper solid and short--dotted curves and symbols $\ast$) and $(p,\gamma)$ reaction (lower similar curves and symbols) on $^{93}$Nb \cite{exfor,fh20}.} 
\end{figure}

{\it The proton OMP} of Koning and Delaroche \cite{KD03} was also the first option for the calculation of the proton transmission coefficients for isotopes of Zr and Nb. 
Nevertheless, it has been checked by analysis of the only available $(p,n)$ reaction data at incident energies of several MeV, as shown for Zr (Figs.~\ref{Fig:Zr9246pgn}) and Nb (Fig.~\ref{Fig:Nb93pgn}).  
The proton OMP fully constrains the calculated $(p,n)$ cross--sections at energies higher than 3--4 MeV, where this reaction channel becomes dominant,  with cross--sections close to optical--potential $\sigma_R$. 

Moreover, the same is true for the $(p,\gamma)$ reaction below the $(p,n)$ reaction effective threshold, where its cross--sections are, in their turn, close to $\sigma_R$ values. 
Thus, the good agreement shown in Fig.~\ref{Fig:Zr9246pgn}(a) between the results within this work and recently measured $^{92}$Zr$(p,\gamma)^{93}$Nb reaction cross--sections at incident energies of 2--5 MeV \cite{as13} supports the global OMP parameters \cite{KD03} even if the earlier $(p,n)$ related data \cite{exfor} are somehow underpredicted between 3 and 4 MeV [Fig.~\ref{Fig:Zr9246pgn}(b)]. 

%subsection{$\gamma$-ray strength functions} \label{RSF}

{\it The radiative strength functions} (RSF) for Nb and especially Mo isotopes do not yet have a confident parametrization despite the widespread systematics \cite{tk20,sg19} and extensive studies (\cite{fh21} and Refs. therein) performed since our former analysis for Zr stable isotopes \cite{va17}. 
More recent accurate cross--section measurements of reactions $^{93}$Nb$(p,\gamma)^{94}$Mo  \cite{fh20} and $(\alpha,\gamma)$ on Zr isotopes \cite{rk21,sh18}, on the other hand, motivated a further survey of these RSFs. 

\begin{figure*} %[h]
\resizebox{1.94\columnwidth}{!}{\includegraphics{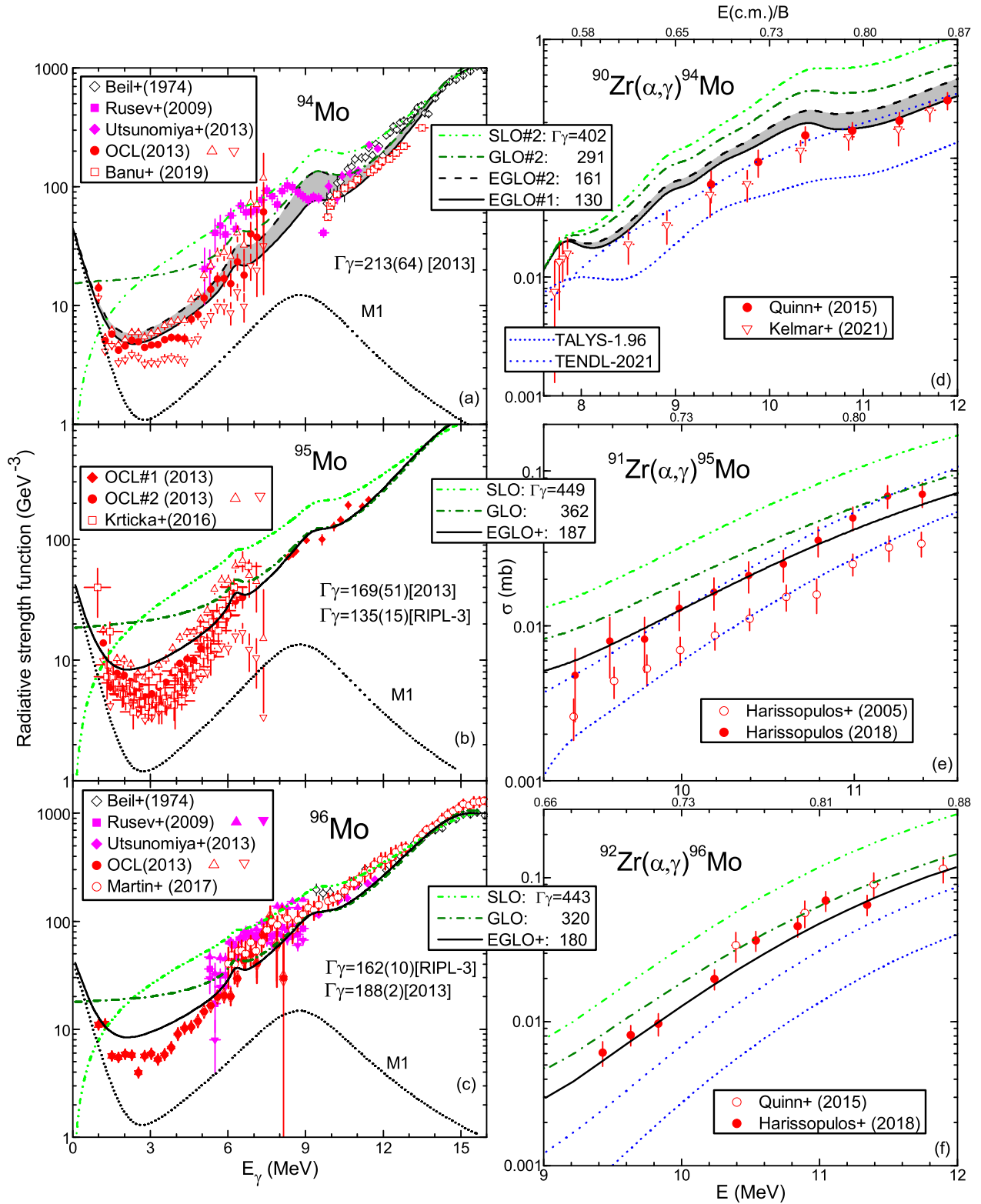}}
\caption{\label{Fig:RSFag-Mo9456} Comparison of: (a--c) measured RSFs of $^{94,95,96}$Mo nuclei \cite{sg19,OCL} and calculated sum of $M$1-radiation SLO model and either $E$1-radiation SLO (dash-dot-dotted), GLO (dash-dotted) RSFs, or the one more sum of upbend--including $M$1 component (short dotted) and EGLO values of (a) either lower (solid) or upper (dashed) resonance parameters for $^{92}$Mo nucleus (Table II \cite{gmt16}), and (b,c) similar middle resonance parameters (solid), along with average $s$-wave  radiation widths $\Gamma_{\gamma}$ (in meV) measured \cite{ripl3,hu13} and calculated as well; (d--f) $^{90-92}$Zr$(\alpha,\gamma)^{94-96}$Mo reaction cross--sections measured \cite{exfor,rk21,sh18}, evaluated \cite{TENDL} (short dashed), calculated by default options of TALYS-1.96 \cite{TALYS} (short-dotted), and with the above--mentioned RSFs (similar curves), vs $\alpha$-particle laboratory energy (bottom) and ratio of center-of-mass energy to Coulomb barrier $B$ \cite{wn80} (top).} 
\end{figure*}

Thus, presently we have adopted the giant dipole resonance (GDR) and additional $M$1 upbend parameters found recently to describe the RSF data for $^{92,94}$Mo nuclei \cite{gmt16}. 
The lower and upper resonance parameters given in Table II of Ref. \cite{gmt16}) provide an RSF uncertainty band for $^{94}$Mo, while the related middle resonance parameters were used for the RSF account for $^{95,96}$Mo nuclei.  
A comparison of the results obtained by using these parameters within the former Lorentzian (SLO) \cite{pa62}, generalized Lorentzian (GLO) \cite{jk90}, and enhanced generalized Lorentzian (EGLO) \cite{jk93} models for the electric-dipole RSF is shown in Figs.~\ref{Fig:RSFag-Mo9456}(a--c). 
The EGLO model has led to an enhanced description of both RSF \cite{OCL,sg19} and average $s$-wave radiation widths $\Gamma_{\gamma}$ \cite{ripl3,hu13} measured data. 
The SLO and GLO models resulted in different RSF energy dependence below the neutron binding energy, as well as larger $\Gamma_{\gamma}$--values.  

Similar results have been obtained for $^{94}$Nb nucleus by using the GDR parameters of Kopecky and Uhl \cite{jk90} but the middle nuclear temperature and $M$1--upbend parameters of $^{92,94}$Mo nuclei \cite{gmt16}. 
The same parameters have been used for $^{93}$Nb, too, with the good agreement shown in Fig.~\ref{Fig:Zr9246pgn}(a) for the EGLO model, at once with $(p,n)$ analysis related to the proton OMP setup. 

Furthermore, a comparative analysis of $^{93}$Nb$(p,\gamma)^{94}$Mo and  $^{90}$Zr$(\alpha,\gamma)^{94}$Mo reactions modeling at low incident energies concerns also the recent data of Heim {\it et al.} \cite{fh20} (Fig.~\ref{Fig:Nb93pgn}) and Kelmar {\it et al.} \cite{rk21} [Fig.~\ref{Fig:RSFag-Mo9456}(d)], respectively. 
A better agreement with these most accurate measured cross--sections corresponds also to the EGLO model for the electric-dipole RSF, versus the related GLO and SLO models. 
There are, however, rather distinct details for the two reactions. 
Thus, the $(p,\gamma)$ cross--sections are described by taking into account both limits of the resonance parameters of Ref. \cite{gmt16}, while only the lower limit is yet close to the $(\alpha,\gamma)$ cross--sections. 
One may conclude that different RSFs could describe the two decay channels of the compound nucleus $^{94}$Mo. 

On the other hand, there are other distinct issues with these reactions. 
So, the angular--momentum ranges of the CN initial population are distinct due to the different target--nuclei g.s., i.e. (9/2$^+$) of the odd $^{93}$Nb and 0$^+$ of the even--even $^{90}$Zr. 
More comments that would be well deserved in this respect  are not, however, the goal of this work. 
Therefore we should note only that the alternate SLO and GLO models, already proven in Fig.~\ref{Fig:RSFag-Mo9456}(a)  to overestimate the RSF data and average $s$-wave  radiation widths, are also leading to much larger cross--sections of both $(p,\gamma)$ and  $(\alpha,\gamma)$ cross--sections. 

Moreover, the same is the case of the EGLO model for $^{95,96}$Mo nuclei and $(\alpha,\gamma)$ reaction on $^{91,92}$Zr, with increased RSFs due to account of the middle resonance parameters \cite{gmt16} and the more recent data of Krti\v cka {\it et al.} \cite{mk16} shown in Fig.~\ref{Fig:RSFag-Mo9456}(b) for $^{95}$Mo. 
Their values, which are higher than the former ones \cite{OCL}, have been taken into consideration for both $^{95,96}$Mo nuclei. 
Consequently, the corresponding $(\alpha,\gamma)$ calculated cross--sections [Figs.~\ref{Fig:RSFag-Mo9456}(e,f)] are increased also for the odd residual nucleus $^{95}$Mo.

%\subsection{Pre-equilibrium emission modeling} \label{PE}

{\it The PE Geometry-Dependent Hybrid (GDH) model} \cite{mb83} was also used, which was generalized by including the angular-momentum and parity conservation \cite{ma90} and knockout $\alpha$-particle emission based on a pre-formation probability $\varphi$ \cite{eg92} . 
It also includes a revised version of the advanced particle-hole level densities (PLD) \cite{ma98,ah98} with a Fermi--gas energy dependence of the single--particle level (s.p.l) density \cite{ck85}. 
The $\alpha$-particle s.p.l. density $g_{\alpha}$=($A$/10.36) MeV$^{-1}$ \cite{eg81}, on the other hand, has been replaced by the value related to the level-density parameter $a$ via the usual equidistant spacing--model relation $g$=(6/$\pi^2$)$a$. 
Moreover, the above--mentioned OMP parameters have also been involved  in the local density approximation (\cite{mb83} and Refs. therein), as also the local--density Fermi energies for various partial waves, corresponding to the central--well Fermi energy value $F$=40 MeV. 

\subsection{Direct reaction account} \label{DR}

Similar to the previous work on $A$$\sim$60 nuclei \cite{va21,va22}, the pickup contributions to $(n,\alpha)$ reactions have been determined within the DWBA method using the code FRESCO \cite{FRESCO}. 
Thus, an one--step reaction has also been considered through the pickup of $^3$He cluster while the "spectator model" \cite{smits76,smits79} was involved. 
The two transferred protons in $(n,\alpha)$ reactions are assumed to be coupled to zero angular momentum, acting as spectators, while the transferred orbital ($L$) and total ($J$) angular momenta are given by the third unpaired neutron of the transferred cluster. 

The prior distorted--wave transition amplitudes and the finite--range interaction were considered, with the $n$-$^3$He and $p$-$t$ effective interactions in the $\alpha$ particle assumed to have a Gaussian shape \cite{eg86,eg88} set by the fit of the binding energies of $^3$He and $t$, respectively. 
Moreover, the bound states of the three--nucleon transferred cluster were generated in a Woods--Saxon real potential \cite{smits79,eg86,eg88} with the depth adjusted to fit the separation energies in the target nuclei. 
The harmonic--oscillator energy conservation rule \cite{smits76,smits79} and the $n$ and $l$ single--particle shell--model state quantum numbers were used to determine the  number of $N$ nodes in the radial three--nucleon cluster wave function. 

Once more, the lack of measured $\alpha$-particle angular distribution, for the $(n,\alpha)$ reactions within this work, made possible only DWBA calculations of related pickup cross--sections using (i) the spectroscopic factors (SFs) of Glendenning (Table II of Ref. \cite{glendenning}) for the spectator proton pair \cite{eg86,eg88,smits79}, in addition to (ii) SFs for the picked neutron that becomes thus responsible for the angular-momentum transfer. 
The latter have been obtained through analysis of $\alpha$-particle angular distributions of one--nucleon pickup reactions $(^3He,\alpha)$ or $(t,\alpha)$ toward the residual nucleus of interest, as shown in the following. 

\begin{figure*} %[h]
\resizebox{1.6\columnwidth}{!}{\includegraphics{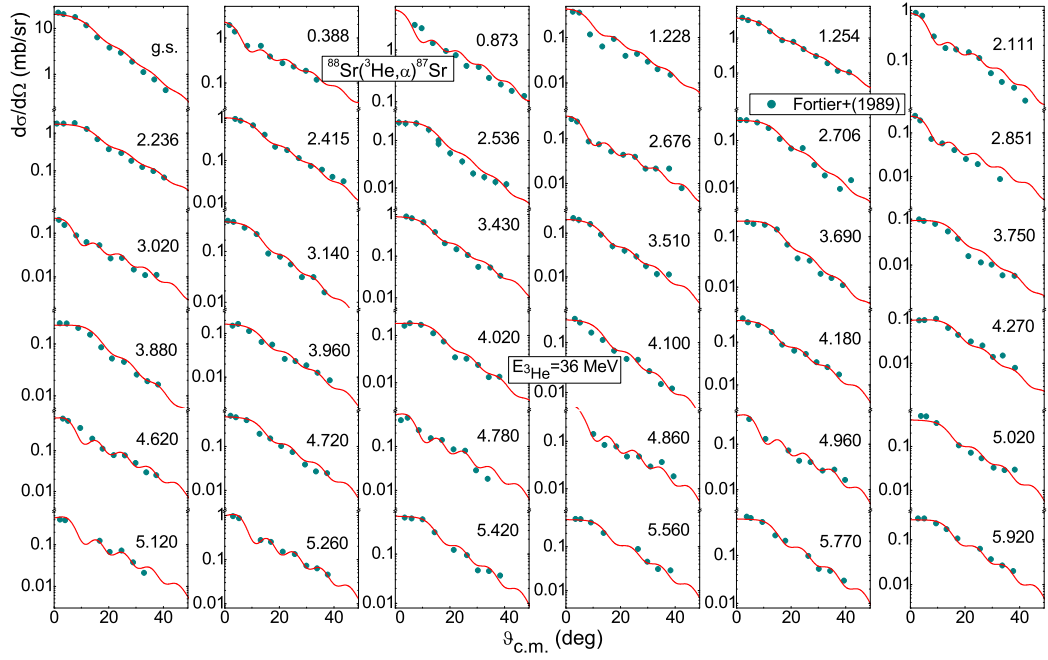}}
\caption{\label{Fig:Sr883HeaSr87} (Color online) Comparison of the measured (solid circles) \cite{sf89} and calculated $\alpha$-particle angular distributions (solid  curves) of $^{88}$Sr$(^3He,\alpha)^{87}$Sr pickup transitions to states shown with excitation energies in MeV, at the incident energy of 36 MeV.}
\end{figure*}

{\it $^{88}$Sr$(^3He,\alpha)^{87}$Sr} $\alpha$-particle angular--distribution \cite{sf89} analysis, at 36 MeV incident energy, provided the SFs of neutrons picked from the $2p$, $2d$, $1f$, $1g$ shells. 
The comparison of measured data and the DWBA calculated angular distribution is shown in Fig.~\ref{Fig:Sr883HeaSr87}. 
A number of 43 levels up to $\sim$6 MeV excitation energy  of $^{87}$Sr  residual nucleus has been considered in this respect, as well as for calculation of $^{90}$Zr$(n,\alpha)^{87}$Sr pickup excitation function. 
The spectator proton pair picked from 2p$\frac{1}{2}$ subshell \cite{eg86,eg88} has been considered. 

\begin{figure} %[h]
\resizebox{1.0\columnwidth}{!}{\includegraphics{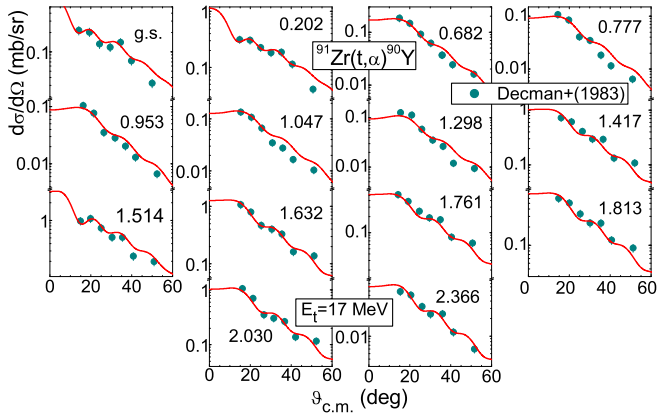}}
\caption{\label{Fig:Zr91taY90} As Fig.~\ref{Fig:Sr883HeaSr87} but for $^{91}$Zr$(t,\alpha)^{90}$Y reaction and incident energy of 17 MeV  \cite{djd83}.}
\end{figure}

\begin{figure} %[h]
\resizebox{1.0\columnwidth}{!}{\includegraphics{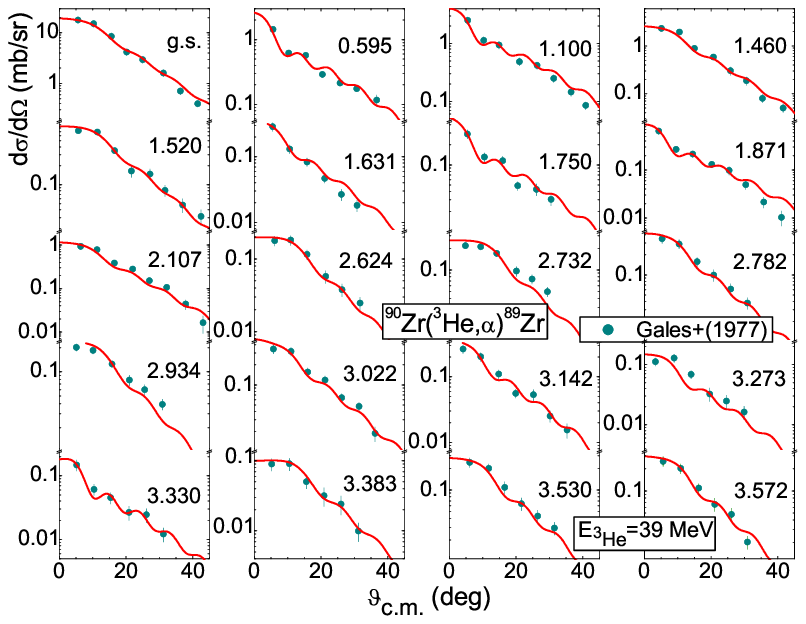}}
\caption{\label{Fig:Zr903HeaZr89} As Fig.~\ref{Fig:Sr883HeaSr87} but for $^{90}$Zr target nucleus and  incident energy of 39 MeV  \cite{sg77Zr90}.}
\end{figure}

\begin{figure} %[h]
\resizebox{1.0\columnwidth}{!}{\includegraphics{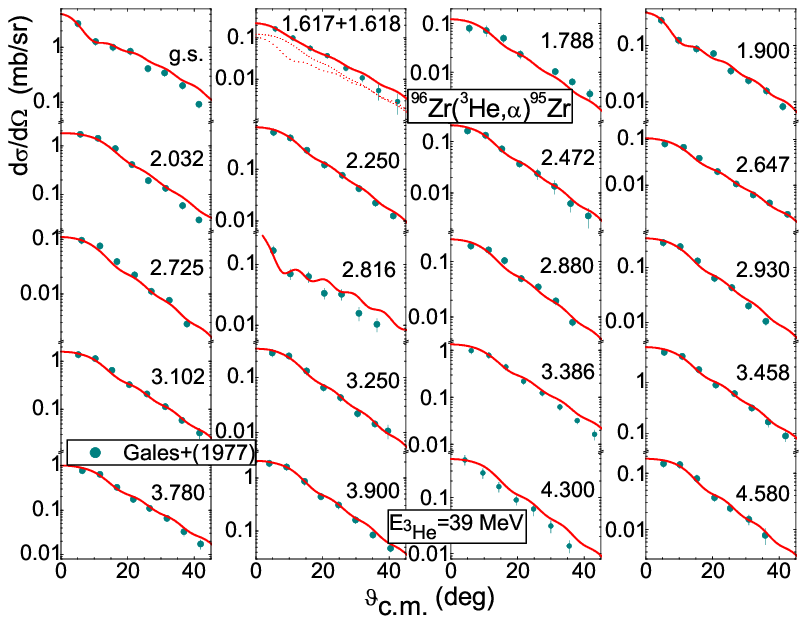}}
\caption{\label{Fig:Zr963HeaZr95} As Fig.~\ref{Fig:Sr883HeaSr87} but for $^{96}$Zr target nucleus and  incident energy of 39 MeV  \cite{sg77Zr96}.}
\end{figure}

{\it $^{91}$Zr$(t,\alpha)^{90}$Y} pickup--reaction analysis and the assumption of a similarity between the picked--proton SFs and picked--neutron SFs \cite{sharp} have been involved for calculation of $^{93}$Nb$(n,\alpha)^{90}$Y reaction cross--sections. 
Thus, 14 levels up to $\sim$ 2.5 MeV excitation energy have been considered within the analysis of the measured $\alpha$-particle angular distributions \cite{djd83} shown in Fig.~\ref{Fig:Zr91taY90}. 
The same excited levels have been concerned with the calculation of the above--mentioned $(n,\alpha)$ pickup excitation function in the following. 

{\it $^{90}$Zr$(^3He,\alpha)^{89}$Zr} reaction angular--distribution \cite{sg77Zr90} analysis, at 39 MeV incident energy, provided the SFs of the picked neutron from the $2p$, $2d$, $1f$, $1g$ shells, too. 
The spectator proton pair picked from 2p$\frac{1}{2}$ subshell has also been considered.  
20 levels of $^{89}$Zr residual nucleus up to 3.572 excitation energy were considered for this reaction (Fig.~\ref{Fig:Zr903HeaZr89}) as well as for assessment of $^{92}$Mo$(n,\alpha)^{89}$Zr pickup  excitation--function calculation of interest for this work. 

{\it  $^{96}$Zr$(^3He,\alpha)^{95}$Zr} $\alpha$-particle angular--distribution analysis, of the measured data also at 39 MeV incident energy \cite{sg77Zr96}, provided the picked--neutron SFs to be used together with Glendenning's SF of the transferred spectator proton pair from 2p$\frac{1}{2}$ subshell. 
A number of 23 levels of $^{95}$Zr up to the excitation energy of 4.58 MeV, involving 2p, 2d, 1f and 1g shells have been considered in this respect  (Fig.~\ref{Fig:Zr963HeaZr95}) as well as, finally, for $^{98}$Mo$(n,\alpha)^{95}$Zr pickup excitation--function calculation.

\section{Results and discussion} \label{Res}

\subsection{Recent $\alpha$-induced reaction data analysis} \label{Resa}

The recently measured cross--sections of $(\alpha,\gamma)$ \cite{rk21,sh18} and $(\alpha,n)$ \cite{ggk21,gh22,tns21,wjo22,mc22} reactions, the former being already taken into account for RSF endorsement, are of particular interest for a customary further validation of the $\alpha$-particle OMP \cite{va14}, too.

\subsubsection{$^{90,91,92}$Zr$(\alpha,\gamma)$$^{94,95,96}$Mo} \label{Resag}

The $^{90}$Zr$(\alpha,\gamma)^{94}$Mo cross--sections \cite{rk21} extended the data energy range below that of the previous experiment \cite{sjq15} taken into account within our former analysis \cite{va17}. 
This extension has touched even the energies around the $(\alpha,n)$ reaction threshold, where $(\alpha,\gamma)$ cross--section is near the $\alpha$-particle total reaction cross--section. 
As a result, the former becomes a strong constraint on the $\alpha$-particle OMP, whereas the RSF has no effect on these calculated reaction cross--section. 
Therefore, the good agreement of calculated and measured data at these energies in Fig.~\ref{Fig:RSFag-Mo9456}(d) supports the $\alpha$-particle OMP \cite{va14} as well. 

A similar extension to lower incident energies was performed for $\alpha$ particles incident on $^{92}$Zr \cite{sh18}, the results shown in Fig.~\ref{Fig:RSFag-Mo9456}(f) providing additional support  for the $\alpha$-particle OMP \cite{va14}. 
Furthermore, the agreement in Fig.~\ref{Fig:RSFag-Mo9456}(e) of presently calculated and increased new measured data for the target nucleus $^{91}$Zr \cite{sh18} has confirmed this $\alpha$-particle OMP, too.

%\subsubsection{$(\alpha,n)$ reaction on $^{96}$Zr and $^{100}$Mo} \label{Resan}
\subsubsection{$^{96}$Zr$(\alpha,n)^{99}$Mo} \label{Resan1}

\begin{figure} %[h]
\resizebox{1.0\columnwidth}{!}{\includegraphics{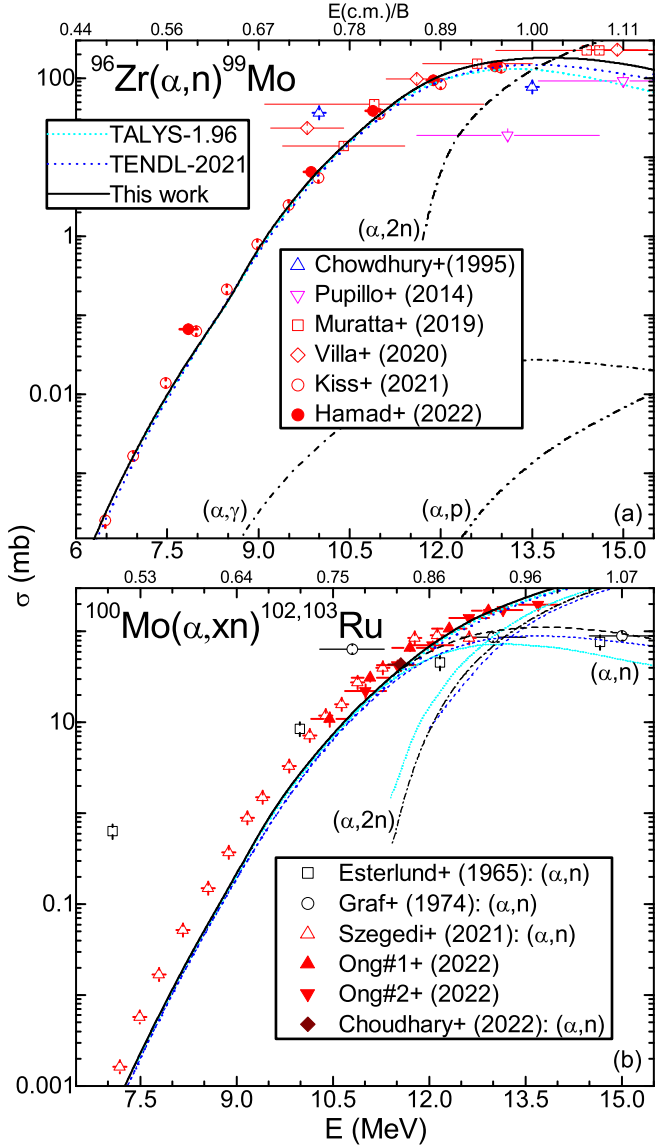}}
\caption{\label{Fig:Zr96Mo100an} Comparison of cross--sections measured \cite{exfor,ggk21,gh22,tns21,wjo22,mc22,tm19,nev20}, evaluated \cite{TENDL} (short--dashed), calculated by default options of TALYS-1.96 \cite{TALYS} (short--dotted), and in this work for $(\alpha,n)$ reaction on (a) $^{96}$Zr (solid curve), and (b) $^{100}$Mo (dashed); there are also shown calculated cross--sections for reactions $(\alpha,2n)$ (dash--dotted), (a) $(\alpha,\gamma)$ (short-dash-dotted) and $(\alpha,p)$ (dash-dot-dotted), and (b) $(\alpha,xn)$ (solid).}
\end{figure}

The first cross--section measurement of $^{96}$Zr$(\alpha,n)^{99}$Mo reaction at incident energies ranging from 6.5 to 13 MeV \cite{ggk21}, more than six orders of magnitude, was also a stringent test of the $\alpha$-particle OMP \cite{va14}. 
Another recent measurement of the same reaction \cite{gh22} supports this issue, with some discrepancies remaining at the lowest incident energy around 8 MeV.  
The use of the present consistent input parameters led to a suitable agreement, i.e. within measurement error bars, with these new data except for an energy range of $\sim$1 MeV around the 8.5 MeV incident energy, as shown in Fig.~\ref{Fig:Zr96Mo100an}(a). 
Calculated cross--sections of the eventually competitive reaction channels are also shown in the same figure, with none of them being able to motivate the lower calculated $(\alpha,n)$ cross--sections at these energies. 
Furthermore, the $(\alpha,n)$ reaction cross--sections account for $\sim$99\% of $\sigma_R$ below the $(\alpha,2n)$ reaction threshold, implying that this problem seems to be related entirely to the $\alpha$-particle OMP.

As a result, the  high accuracy of the new data \cite{ggk21,gh22} at these energies is  critical for an additional assessment of the surface imaginary--potential depth $W_D$, which increases between the energies $E_1$ and $E_2$ (Table II and Fig. 1 of Ref. \cite{va14}) due to a significant change in the number of open reaction channels close to the Coulomb barrier $B$ \cite{ma09,ma10,pm97}.  
This $W_D$ trend below an energy $E_2$ corresponding to 0.9$B$ \cite{ma09,ma10} has been proved by reaction cross--section analysis \cite{va14,ma09,ma10} while the elastic--scattering studies have shown that $W_D$ decreases at higher energies. 

Because $E_1$=8.55 MeV and $E_2$=12.15 MeV for $\alpha$ particles incident on $^{96}$Zr, the lower calculated $(\alpha,n)$ cross--sections near $E_1$ have an immediate significance. 
Thus, even at lower incident energies, the appropriate data account at even lower incident energies has revealed that the related constant minimum $W_D$ has a proper value but the sudden change at $E_1$ is less physical (as expected for a model). 
The agreement of the measured and calculated cross--section at energies slightly above this limit supports the $W_D$ energy dependence within its most important range \cite{ma09,ma10}. 
An eventual smoothing of the $W_D$ abrupt change at $E_1$ may significantly improve the data underestimation around this energy.  

\subsubsection{$^{100}$Mo$(\alpha,xn)^{102,103}$Ru} \label{Resan2}

A recent measurement of $^{100}$Mo$(\alpha,n)^{103}$Ru reaction cross--sections \cite{tns21} at low energies based on the difference in activation thick target yields at two neighboring energies is also available. 
Because no agreement was found with previous measurements within their reported uncertainties [Fig.~\ref{Fig:Zr96Mo100an}(b)], the same target nucleus was chosen as a proof-of-principle measurement in the inverse kinematics of $(\alpha,xn)$ inclusive cross--sections between 8.9--13.2 MeV in the center of mass \cite{wjo22}. 
Moreover, while new $(\alpha,n)$ cross--sections have been reported in the energy range 11--32 MeV \cite{mc22}, too, only the lowest energy point has been considered hereafter to avoid additional PE effects within this discussion. 

Except for the highest energy, the $(\alpha,xn)$ cross--sections calculated in this work using the $\alpha$-particle OMP \cite{va14} agree with the most recent measured data \cite{wjo22,mc22} within their error bars  [Fig.~\ref{Fig:Zr96Mo100an}(b)]. 
%However, one may note that the lower value of this data point would be physical in the case of an excitation--function plateau, which is not the case of this $(\alpha,xn)$ reaction. 
Except for, once again, the highest energy of this measurement following a sharp change in these data slope, the $(\alpha,n)$ activation cross--sections of Ref. \cite{tns21} are significantly underestimated. 
The disagreement's energy range includes the two above--mentioned energy limits of our OMP, which in this case are $E_1$=9.08 MeV and $E_2$=12.61 MeV \cite{va14}. 
However, there is now a constant increase in the calculated $(\alpha,n)$ cross--sections over $E_1$ as well as a suitable agreement between experimental and calculated results around $E_2$. 
One may also note the close similarity between the two neutron--rich nuclei $^{96}$Zr and $^{100}$Mo, as heaviest natural isotopes of their elements and similar nuclear asymmetry $(N-Z)/A$. 
Thus, the distinct effects on calculated cross--sections at energies below $B$, where $W_D$ increases with $\alpha$-particle energy, should be investigated further. 

Nonetheless, the newer data are already well described by the $\alpha$-particle OMP \cite{va14} around the energy limit $E_2$, i.e. around the surface imaginary--potential depth maximum. 
The relevance of this energy below which the potential parameters have to be strongly modified was also considered by Sauerwein {\it et al.} \cite{as11}, based on Ref. \cite{ma10}, as well as more recently \cite{nnl22,db22}.  
Their correction by a further Fermi--type function, on the other hand, did not concern the surface but volume imaginary--potential depth $W_V$ of the indeed much simpler OMP of McFadden and Satchler \cite{lmf66}. 
This may account for the quite different values obtained for the 'diffuseness' of this Fermi--type function. 

\begin{figure*} %[h]
\resizebox{1.94\columnwidth}{!}{\includegraphics{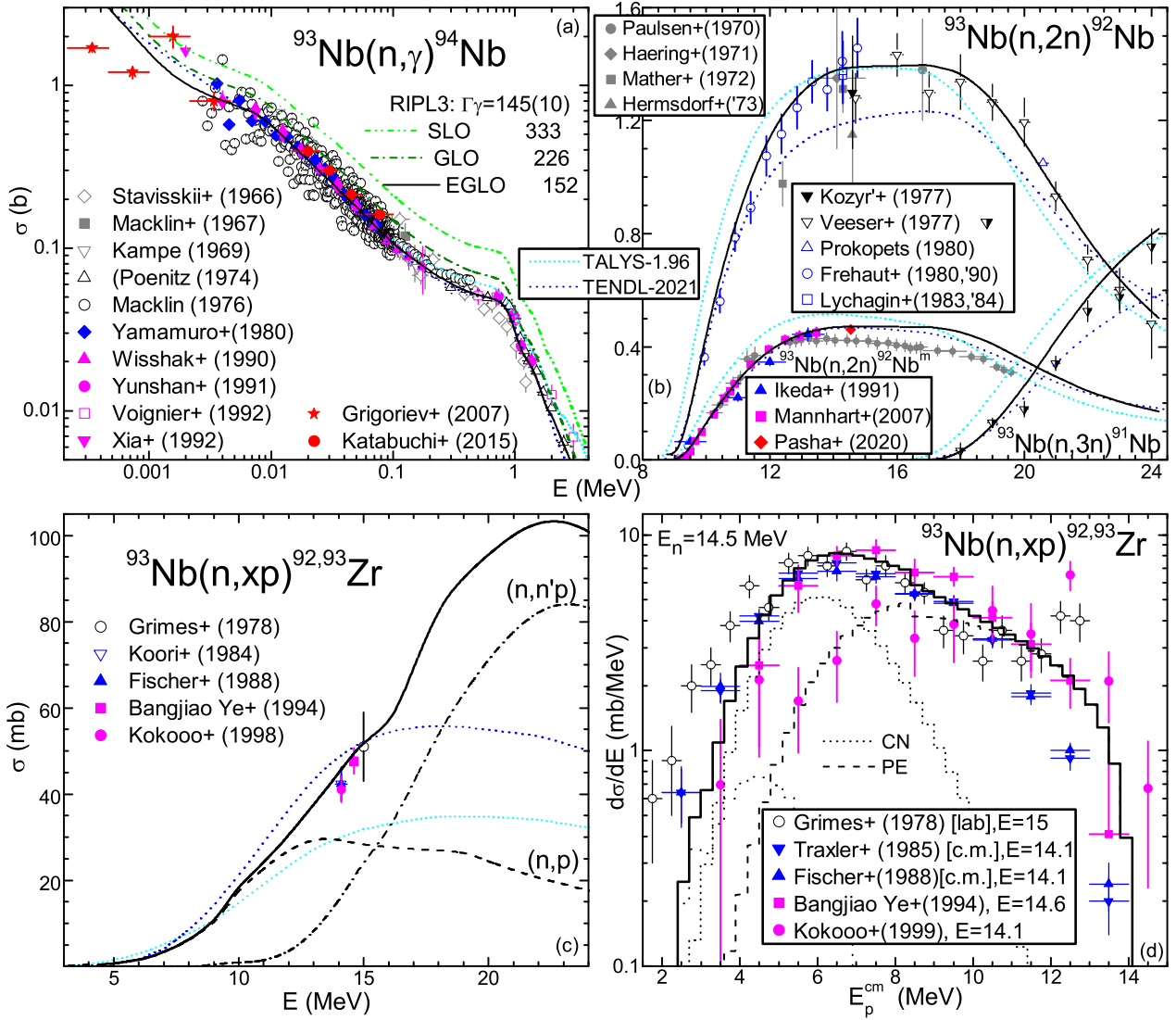}}
\caption{\label{Fig:Nb93nx} (Color online) As Fig.~\ref{Fig:Zr9246pgn}, of calculated cross--sections in this work (solid curves), but for (a-c) the $(n,\gamma)$, $(n,2n)$, and $(n,xp)$ reactions on $^{93}$Nb, and (d) 14.5 MeV neutron--induced proton spectrum \cite{smg78,gt85,rf88,by94,k99} with $(n,p)$ (short-dotted) and $(n,n'p)$ (dotted) CN, and PE (dashed) components.}
\end{figure*}

\subsection{Nucleon--emission induced by neutrons on $^{93}$Nb} \label{Resn1}

There is a large amount of experimental data for neutron interaction with $^{93}$Nb nucleus due to interest in it for structural materials of nuclear reactors, activation monitor in reactor dosimetry, 14 MeV neutron flux determination, and also as an element of superconductor alloys in fusion reactors. 
It triggered consideration of this interaction even as a {\it 'sample problem'} \cite{hg85} in nuclear model calculations (e.g., \cite{eg92,ck00,yw95,pd96,ck98}). 
However, the scattered data for the $(n,\alpha)$ reaction on this single Nb natural isotope suggest that more precise measurements are needed to settle its evaluation (e.g., Refs. \cite{ai16,hl17}). 

Therefore, a prerequisite for a consistent discussion of the $\alpha$-particle emission is a suitable account, by using the actual parameter set, of all competing nucleon-emission data for neutrons incident on $^{93}$Nb. 
The same is the case for $^{92}$Mo, in completion of the earlier study of neutron--induced reactions on Mo stable isotopes \cite{pr05}. 
Upon due consideration of these two nuclei, $\alpha$-emission analysis will be feasible for Zr, Nb, and Mo nuclei altogether. 

{\it The $(n,\gamma)$ cross--section} analysis should be  considered first because of the significant isomeric-states activation by neutrons on $^{93}$Nb. 
Thus, a former validation of the $\gamma$-ray transmission coefficients will constrain the isomeric cross--sections to the adopted NLD spin cutoff factors. 
The particular agreement of the more recent experimental data and the calculated results corresponding to the EGLO model for the electric--dipole RSF is similar to that of the related average $s$-wave radiation widths likewise shown in Fig.~\ref{Fig:Nb93nx}(a). 
The GLO and especially the SLO models provide calculated cross--sections as well as $\Gamma_{\gamma}$ values that are much larger. 

\begin{figure*} %[h]
\resizebox{1.94\columnwidth}{!}{\includegraphics{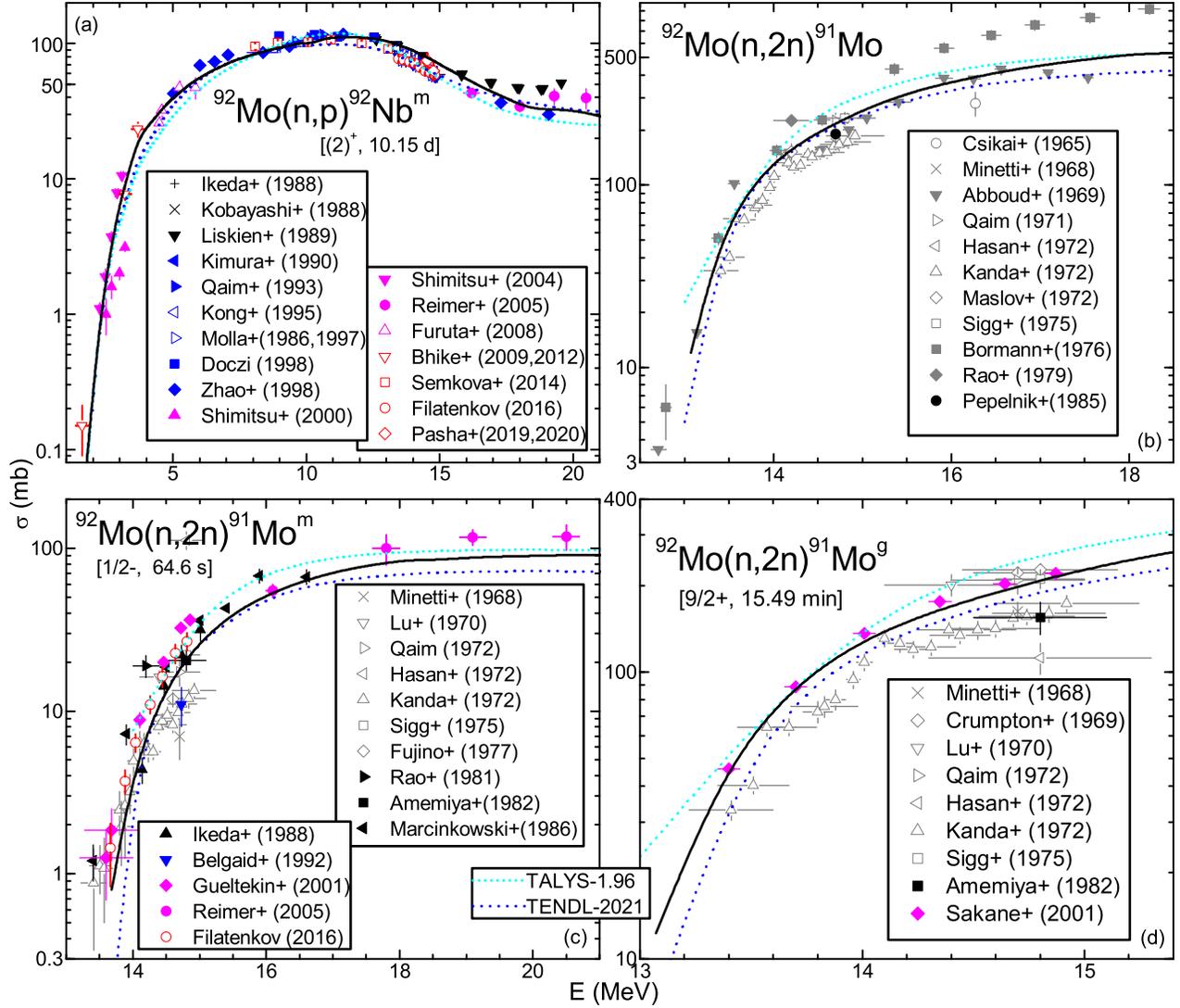}}
\caption{\label{Fig:Mo92nx} (Color online) As Fig.~\ref{Fig:Nb93nx}(b) but for $(n,xp)$ and $(n,2n)$  reactions on $^{92}$Mo \cite{exfor}.}
\end{figure*}

{\it The $(n,2n)$ and $(n,xp)$ reactions} analysis must contend a  lack of total cross--section measurements within the latest 40 years. 
Fortunately, there are recent measurements for the isomeric cross--sections corresponding to the 2$^+$ state of $^{92}$Nb nucleus at 136 keV, as shown in  Fig.~\ref{Fig:Nb93nx}(b). 
Of particular interest is the agreement of the calculated and recently measured cross--section at the incident energy of $\sim$14 MeV, i.e., on the flat maximum of this excitation function. 
In contrast to TALYS-1.96 as well as TENDL-2021 results, this  concurrently appropriate account of this isomeric state and total $(n,2n)$ excitation functions supports both neutron OMP and NLD spin dependence.  

{\it The proton emission} induced by neutrons on $^{93}$Nb was the subject of several angle--integrated energy distribution studies around the incident energy of 14 MeV, but no measurement of its excitation function has been made. 
Nevertheless, the total proton--emission cross--sections corresponding to these data could be considered at once, with the good agreement shown in Fig.~\ref{Fig:Nb93nx}(c) for the calculated results of this work. 
The overall account of the measured energy spectra, in the limit of the error bars [Fig.~\ref{Fig:Nb93nx}(d)], may support an appropriate description of nucleon emission in neutron--induced reactions on $^{93}$Nb.

\subsection{Nucleon--emission induced by neutrons on $^{92}$Mo} \label{Resn2}

The  earlier systematic investigation of neutron--induced reactions on Mo stable isotopes up to 20 MeV \cite{pr05} included a local approach and consistent parameter set requirements quite similar to the present analysis. 
As a result, while different parameters were formerly involved, an analysis of other independent data concerned also their setting up. 
A definitive account of all available data for competitive reaction channels was obtained, too. 

On the other hand, even the actual $\alpha$-emission data correspond merely to four $^{92,95,98,100}$Mo of the seven stable isotopes of molybdenum. 
Furthermore, due to the larger amount of available measured data, a reanalysis of the nucleon-emission has now only concerned the lighter and even--even semi--magic nucleus $^{92}$Mo.  
It matters also the higher charged--particle emission cross--sections owing to the isotopic effect triggered by reaction $Q$-values, i.e., the CN cross--section decreases with the isotope mass increase \cite{nim77}. 

{\it The $(n,p)$ reaction} analysis for the isomeric--state $^{92}$Nb$^m$ population has been improved as a result of additional data published in the interim, as shown in Fig.~\ref{Fig:Mo92nx}(a). 
The better agreement with these data at lower incident energies has thus validated the neutron as well as proton OMPs adopted in the present work in comparison to the local parameter sets of Ref. \cite{KD03}. 

{\it The $(n,2n)$ reaction} cross--sections have no additional measurements concerning both ground and metastable states as well as their population sum [Fig.~\ref{Fig:Mo92nx}(b-d)]. 
An effective point is the location of the presently calculated results in between those of TALYS-1.96 default predictions and TENDL-2021 evaluated values. 
The change from the default results to the final evaluation is obviously approaching the experimental data. 
However, our calculated cross--sections are, e.g. at higher incident energies, near either the default results for the isomeric cross--sections or the evaluated ones for g.s. population. 
Nevertheless, in both cases, they are closer to the more recently measured data.

\subsection{ $\alpha$-emission spectra and excitation functions} \label{Resna}

\subsubsection{Angle--integrated energy spectra analysis} \label{ResnaDE}

\begin{figure} %[h]
\resizebox{1.00\columnwidth}{!}{\includegraphics{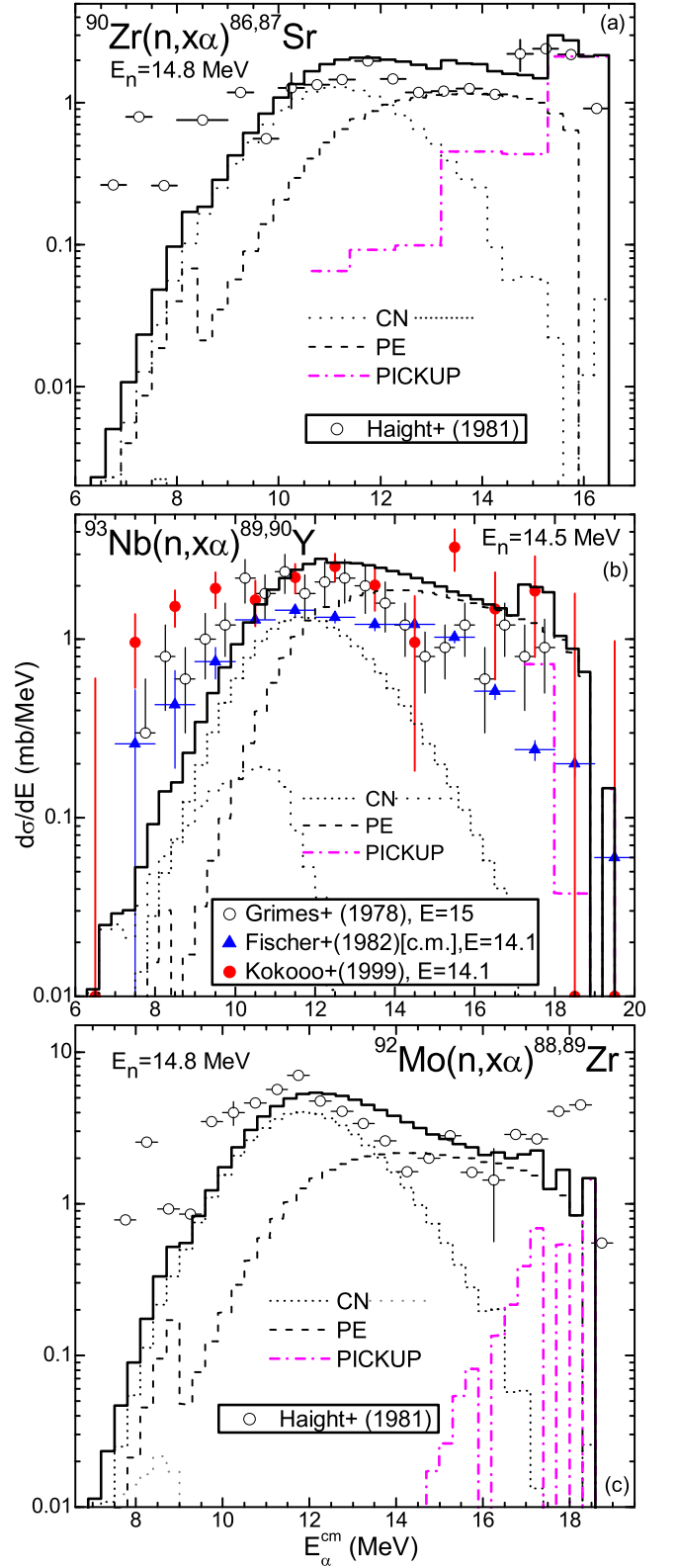}}
\caption{\label{Fig:ZrNbMonaS} (Color online) As Fig.~\ref{Fig:Nb93nx}(d) but for $\sim$14.5 MeV neutron--induced $\alpha$ spectra on (a,c) $^{90}$Zr and $^{92}$Mo \cite{rch81}, and (b) $^{93}$Nb \cite{k99,smg78,rf82}, and DR component (dash-dotted).}
\end{figure}

First, the $\alpha$-particle angle--integrated energy distributions induced by neutrons on  $^{90}$Zr \cite{rch81}, $^{93}$Nb \cite{k99,smg78,rf82}, and $^{92}$Mo \cite{rch81} have been examined to validate the PE component of the energy--spectra above the CN one. 
Thus, the overall account of the measured spectra, in the limit of the error bars for the more accurate data shown in Fig.~\ref{Fig:ZrNbMonaS}, may support an appropriate description of $\alpha$-particle PE emission corresponding to (i) the above--mentioned $\alpha$-particle s.p.l. density $g_{\alpha}$ related to the level-density parameter $a$, and (ii) $\alpha$-particle pre-formation probability $\varphi$ values of 0.1, 0.14. and 0.11 for Zr, Nb, and Mo isotopes, respectively. 
Furthermore, the analysis of these spectra as well as the  $(n,\alpha)$ excitation functions discussed below, suggests a corresponding $\Delta$$\varphi$$\sim$0.02 uncertainty. 

Second, the high--energy limit of these spectra makes possible the check of the DR pickup cross--sections obtained by using the SF for the picked nucleon from the analysis of $\alpha$-particle angular distributions corresponding to one--nucleon pickup reactions $(^3He,\alpha)$ or $(t,\alpha)$, and the spectator proton--pair SF \cite{glendenning} (Sec.~\ref{DR}). 

\begin{figure*} %[h]
\resizebox{1.94\columnwidth}{!}{\includegraphics{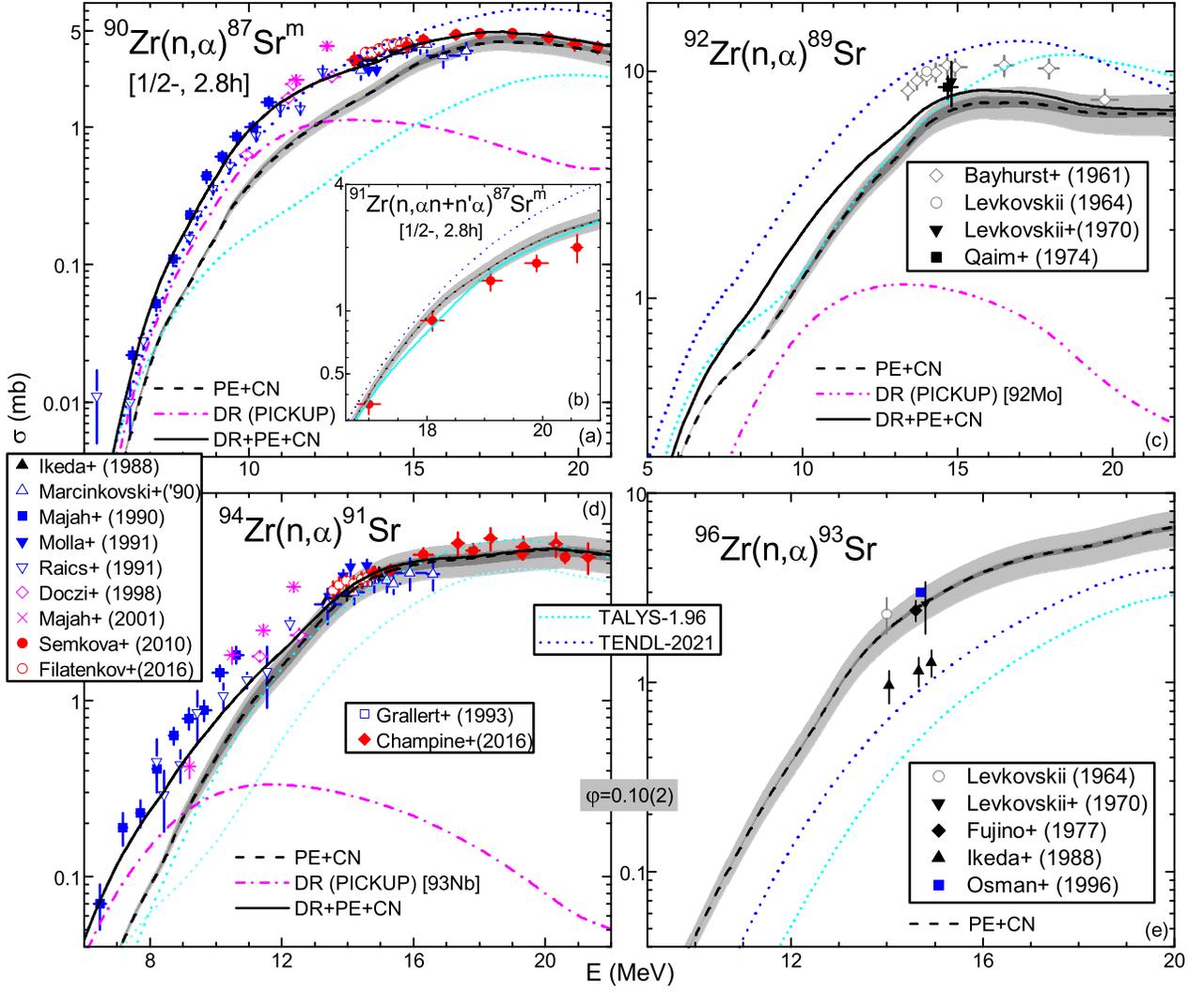}}
\caption{\label{Fig:Zr901246na} (Color online) As Fig.~\ref{Fig:Nb93nx} but for $(n,x\alpha)$ reaction on $^{90,91,92,94,96}$Zr \cite{exfor}, DR (dash-dotted curve) and PE+CN (dashed) components, and PE+CN uncertainty bands related to either residual--nuclei LD (gray) or PE parameter $\varphi$ (light--gray).}
\end{figure*}

The same residual--nucleus levels have been considered within the analysis of both the $\alpha$-particle angular distributions of the one--nucleon pickup reactions $(^3He,\alpha)$ or $(t,\alpha)$, and the present $(n,\alpha)$ pickup excitation functions. 
As a result, the most consistent DR component corresponds to the 43 levels of $^{87}$Sr residual nucleus up to $\sim$6 MeV excitation energy, which is important for calculating the $^{90}$Zr$(n,\alpha)^{87}$Sr spectrum in Fig.~\ref{Fig:ZrNbMonaS}(a). 
Within the highest 2 MeV, it makes a significant contribution to the agreement of calculated and measured spectra. 
However, there is just a minor increase at lower spectrum energies, i.e., higher excitations of the residual nucleus. 

On the other hand, only 14 levels up to $\sim$ 2.5 MeV excitation energy have been considered within analysis of the $\alpha$-particle angular distributions from $^{91}$Zr$(t,x\alpha)^{89,90}$Y, as shown in Fig.~\ref{Fig:Zr91taY90}. 
As a result, a significant DR component within only $\sim$2 MeV at the end of the spectrum of $^{93}$Nb$(n,\alpha)^{90}$Y reaction is shown in Fig.~\ref{Fig:ZrNbMonaS}(b).  

Moreover, the DR component of 20 levels up to 3.572 MeV excitation of $^{89}$Zr residual nucleus, formerly taken into account in $^{90}$Zr$(^3He,\alpha)^{89}$Zr angular--distribution analysis (Fig.~\ref{Fig:Zr903HeaZr89}), is also less important for the $\alpha$-particle energy spectrum of $^{92}$Mo$(n,\alpha)^{89}$Zr reaction [Fig.~\ref{Fig:ZrNbMonaS}(c)].

\subsubsection{Zr isotopes} \label{ResnaZr}

{\it $^{90}$Zr$(n,\alpha)^{87}$Sr$^m$ excitation function} [Fig.~\ref{Fig:Zr901246na}(a)] highlights the importance of the pickup DR component, which is higher than that of CN+PE up to the incident energy of $\sim$12 MeV. 
Then it decreases to an order of magnitude below the latter around 20 MeV. 
Their sum is, however, in good agreement with the experimental data along the whole incident--energy range, while several issues should be underlined.  

First, there is a relevant data account at lower incident energies, where neither NLD nor PE effects exist on the calculated HF cross--sections. 
This is shown in Fig.~\ref{Fig:Zr901246na}(a) by the uncertainty bands corresponding to either the error bars of  $N_d$ and LD parameter $a$ of the residual nucleus $^{87}$Sr (Table~\ref{densp}), or the above--mentioned  $\Delta\varphi$=0.02 incertitude of the main PE parameter. 
Both of them are minimal at these energies and to $\sim$20 MeV due to the low spin of $^{87}$Sr$^m$ isomer, leading to a reduced side-- and cascade--feeding. 
Nonetheless, the absence of other HF+PE uncertainty factors on calculated cross--sections at low energies indicates that the appropriate data account supports the $\alpha$-particle OMP \cite{va14}. 

\begin{figure*} %[b]
\resizebox{1.94\columnwidth}{!}{\includegraphics{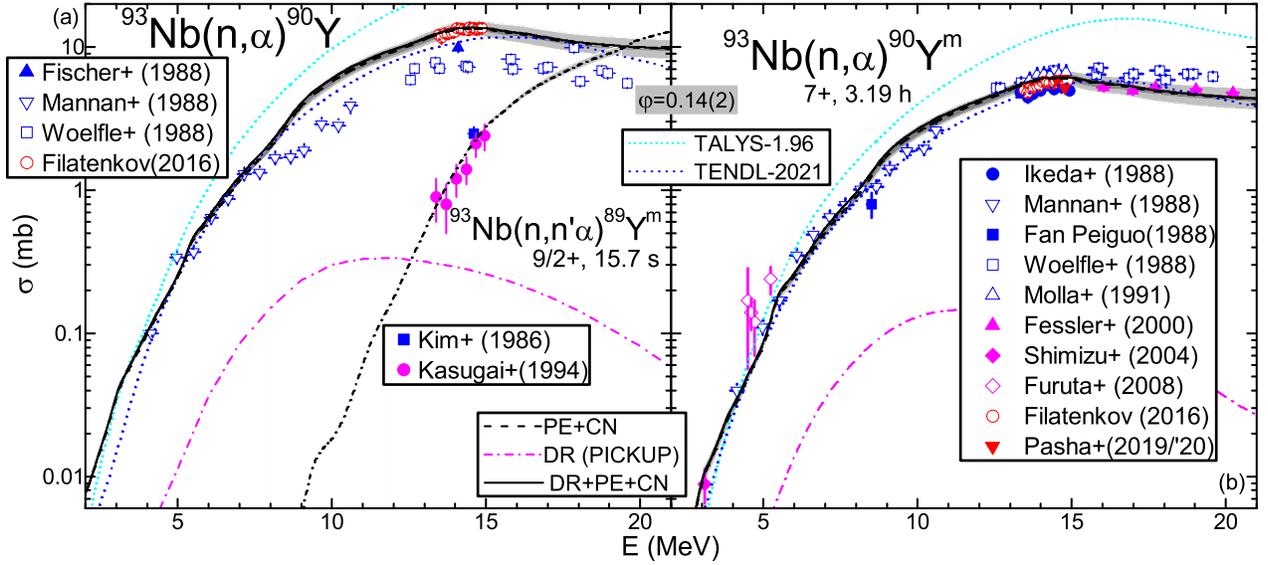}}
\caption{\label{Fig:Nb93na} (Color online) As Fig.~\ref{Fig:Zr901246na} but for the target nucleus  $^{93}$Nb \cite{exfor} and (a) $^{93}$Nb$(n,n'\alpha)^{89}$Y$^m$ reaction (short dash-dotted curve).}
\end{figure*}

The agreement of the measured and calculated cross--sections at higher energies, however, does not follow the already lower DR contribution, but rather an increased PE component in comparison to our previous analysis \cite{va17}. 
It follows the above--mentioned $\alpha$-particle s.p.l. density $g_{\alpha}$ value, proved in the present work by extended energy--spectra analysis and $^{93}$Nb target--nucleus case. 
Nonetheless, the appropriate account of the excitation--function upper side supports both this $g_{\alpha}$ value and the  parameter $\varphi$--value firstly suggested by 14--MeV spectra analysis. 

{\it $^{91}$Zr$(n,n\alpha)^{87}$Sr$^m$ excitation function} [Fig.~\ref{Fig:Zr901246na}(b)] stands as a fine case of the same nucleus in a different reaction channel and $\alpha$-particle energy range. 
The activation of $^{87}$Sr in neutron--induced reactions on $^{91}$Zr nucleus, through $(n,$$\alpha$$n)$ reaction is more than 10 times larger in comparison with that by $(n,n'\alpha)$ reaction, at incident energies of 15-21 MeV. 
%Thus, the $\alpha$-emission energies corresponding to this activation are lower than center--of--mass incident energy by at least the neutron--binding energy within firstly excited $^{88}$Sr nucleus. 
%So, no DR contribution on low-lying levels is anymore possible by $(n,$$\alpha$$n)$ reaction, while only CN processes match the  $(n,n'\alpha)$ reaction. 
Thus, the suitable agreement of measured \cite{vs10} and calculated cross--sections of this reaction, in the absence of DR effects, has  supported the $\alpha$-particle OMP \cite{va14} once more. 
Actually this conclusion has already been reachable within the previous analysis for Zr isotopes \cite{va17}, without DR consideration, which provided a good account only for this reaction. 

{\it $^{92}$Zr$(n,\alpha)^{89}$Sr excitation function} analysis has the drawback of no SF available for the residual nucleus $^{89}$Sr and also no measured data at energies below 14 MeV [Fig.~\ref{Fig:Zr901246na}(c)]. 
Nevertheless, the PE+CN calculated results are just below the more recently measured data, with at least the PE uncertainty band matching their error bars. 
However, to overcome the former shortcoming, we may assume a DR contribution similar to that of $^{92}$Mo$(n,\alpha)^{89}$Zr reaction, mentioned in Sec.~\ref{DR} and in the following. 
Its addition to the CN+PE component provides agreement with both the recently measured cross--sections, in the limit of their error bars, and the excitation--function trend. 
The somehow lower calculated cross--sections are related to the fact that this is merely an attempt to obtain a realistic outline of this $(n,\alpha)$ reaction on $^{92}$Zr. 

{\it $^{94}$Zr$(n,\alpha)^{91}$Sr excitation function} was measured within several experiments, as shown in Fig.~\ref{Fig:Zr901246na}(d), but proper SFs for the residual nucleus $^{91}$Sr are also lacking. 
At the same time, the comparison of the CN+PE component and available data shows a quite different look below and above incident energy of $\sim$14 MeV. 
Thus, quite recent and accurate data between 14--21 MeV are well described, with error bars just within the calculated NDL and PE uncertainty bands, though below 10 MeV there is an underestimation of several times. 
This was one of the major flows in our previous analysis of  neutron--induced $\alpha$-emission on Zr isotopes \cite{va17} for the $\alpha$-particle OMP \cite{va14}. 
However, the assumption of a DR contribution. e.g., for a nearby target nucleus, may  explain these features. 

Thus, a similar DR contribution closer to this case could be that related to the residual nucleus $^{90}$Y in the $(n,\alpha)$ reaction on $^{93}$Nb (Sec.~\ref{DR} and just below). 
Because its maximum value of around 12 MeV is more than an order of magnitude lower than the CN+PE sum, the upper side of $^{94}$Zr$(n,\alpha)^{91}$Sr calculated excitation function  remains unchanged. 
On the other hand, these DR cross--sections are, as for the target nucleus $^{90}$Zr, higher than the own CN component at the incident energies below 9 MeV. 
Their sum is quite close to the measured data as well as the results obtained previously \cite{va17} by using the $\alpha$-emission OMP \cite{va94}.  

{\it $^{96}$Zr$(n,\alpha)^{93}$Sr excitation function} [Fig.~\ref{Fig:Zr901246na}(e)] has the same attributes as those for the target nucleus $^{92}$Zr. 
The calculated CN+PE component and its PE uncertainty band describe rather well the trend of all measured data as well as cross--section values around 14 MeV except for one disparate data set. 
The difference from previous analysis \cite{va17} is related again to the above--mentioned PE contribution change due to the use of the $\alpha$-particle s.p.l. density $g_{\alpha}$ value. 
No further effect of an eventual DR contribution may be considered, the expected outcome being the same as for $^{94}$Zr.

\subsubsection{$^{93}$Nb target nucleus} \label{ResnaNb}

The DR component, which is shown at the high--energy end of the $\alpha$-particle energy spectrum in Fig.~\ref{Fig:ZrNbMonaS}(b), may explain the minor pickup DR contribution to $^{93}$Nb$(n,\alpha)^{90}$Y reaction total cross--sections [Fig.~\ref{Fig:Nb93na}(a)]. 
The same is true for the 7$^+$ isomer of the residual nucleus $^{90}$Y which was also recently measured as shown in Fig.~\ref{Fig:Nb93na}(b) for energies from the effective threshold to above 20 MeV. 
Thus, the agreement of measured and calculated cross--sections stands for CN+PE results, with the main PE uncertainty band just across the error bars of the recent data. 

The latest comment, concerning the 7$+$ isomeric state $^{90}$Y$^m$ activation, is fully appropriate to the activation of the 9/2$^+$ isomer via $^{93}$Nb$(n,n'\alpha)^{89}$Y$^m$ reaction. 
The PE uncertainty band corresponding to this reaction has pointed out no PE effects, too. 
Therefore the good account of all experimental data does validate entirely the CN+PE parameters that matter, namely the $\alpha$-particle OMP \cite{va14}. 

\subsubsection{Mo isotopes} \label{ResnaMo}

\begin{figure*} %[h]
\resizebox{1.94\columnwidth}{!}{\includegraphics{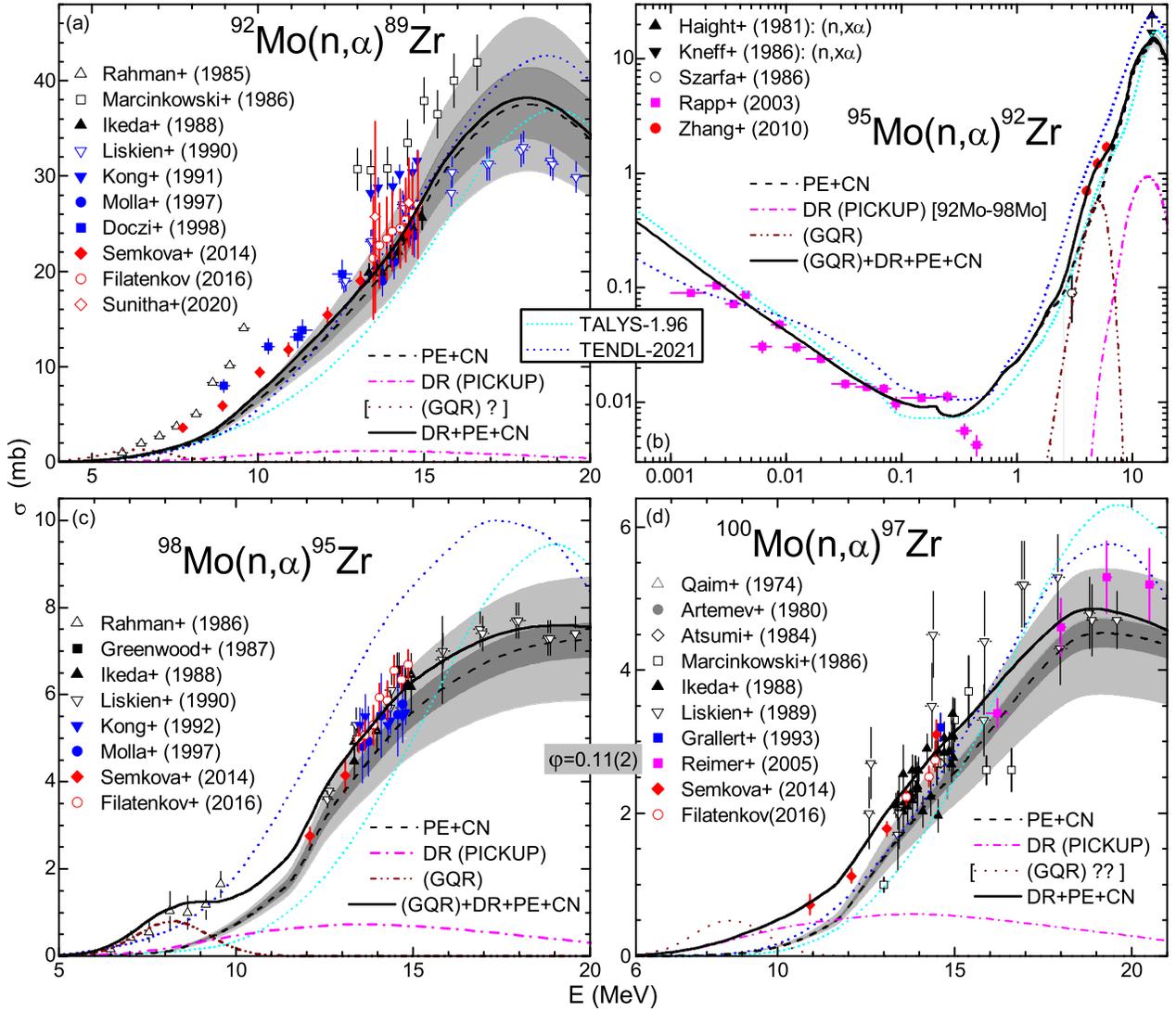}}
\caption{\label{Fig:Mo92580na} (Color online) As Fig.~\ref{Fig:Zr901246na} but for the target nuclei  $^{92,95,98,100}$Mo \cite{exfor} and {\it like}--GQR (dash-dot-dotted) eventual components.}
\end{figure*}
 
{\it $^{92}$Mo$(n,\alpha)^{89}$Zr excitation function} analysis reveals a notable balancing of the DR and PE+CN mechanisms for neutron--induced $\alpha$-emission on light isotopes of elements. 
Thus, the former brings about $\sim$1 mb around 14 MeV, for SFs discussed in Sec.~\ref{DR}, while the measured cross--sections amount to several tens of mb \cite{nim77}. 
Hence, the DR addition to the CN+PE component shown in Fig.~\ref{Fig:Mo92580na}(a) is not significant. 
Therefore, the comparison of experimental and calculated cross--sections of this reaction is related first and foremost to HF+PE modeling. 

However, due to the DR component, there is a clear agreement between calculated and recently measured cross--sections between 12--15 MeV. 
After that, there are quite different cases at higher and lower energies. 
Thus, there are only two disparate data sets available at higher energies, even beyond the LD uncertainty band but within the larger similar PE band. 
On the other hand, at lower energies there is an underestimation up to a factor of two around the incident energy of 8 MeV. 
Meanwhile, the uncertainty bands, which are yet dropped at these energies in Fig.~\ref{Fig:Mo92580na}(a), no longer exhibit LD and PE effects.  

Previously, an apparent increase in measured $\alpha$-emission beyond the DR+PE+CN cross--sections, in neutron--induced reactions on $A$$\sim$60 nuclei \cite{va21,va22}, was found around the GQR energies $E_{GQR}$=$65A^{-1/3}$ MeV \cite{js81} of the related excited nuclei. 
Thus, $\alpha$-particle decay of giant resonances populated via neutron capture has been assumed, with Gaussian distributions added in this respect to the DR+PE+CN sum. 
The widths and peak cross--sections of these distributions were obtained by fitting the extra yields. 
However, because these widths are lower than the systematic 'best' values \cite{js81}, we have called them only {\it like}--GQR components. 
On the other hand, no enhancement beyond the DR+PE+CN cross--sections has been found in the present work for neutrons incident on Zr and Nb isotopes at energies between 5.8--8.7, and 7.1 MeV, respectively, corresponding to these excited--nuclei GQR energies.  
Furthermore, the same energy for the target nucleus $^{92}$Mo is 6.3 MeV, so that an eventual {\it like}--GQR contribution at the actual extra--yield maximum around $\sim$8 MeV would be less significant. 

%Therefore, this extra-yield beyond the DR+PE+CN cross--sections can not be described by assumption of a {\it like}--GQR component corresponding to GQR energy of 14.35 MeV for the excited nucleus $^{93}$Mo. 
Nevertheless, for the sake of discussion, an eventual Gaussian distribution with a peak cross--section of 1 mb at 6.3 MeV related to the GQR energy of 14.35 MeV for the excited nucleus $^{93}$Mo, and a width of 2.35 MeV \cite{va21}, is shown in Fig.~\ref{Fig:Mo92580na}(a). 
The peak cross--section value is chosen to fit the apparent extra-yield at this GQR energy, while the related Gaussian distribution may neither be supported by the available  experimental data nor describe the actual extra-yield around the incident energy of 8 MeV. 
Hence, it has not been included within the finally calculated DR+PE+CN cross--sections. 
Thus, the underestimation of the measured data below 12 MeV remains an open question at variance with the good agreement around 14 MeV. 

{\it $^{95}$Mo$(n,\alpha)^{92}$Zr excitation function} [Fig.~\ref{Fig:Mo92580na}(b)] is a thoroughly different case, not only with an outstanding cross--section set measured from 1 to 500 keV \cite{wr03} but also recent data between 4 and 6 MeV \cite{gz10}. 
The energies of the latter measurement are just around 5.1 MeV incident energy which is linked to the GQR energy of 14.2 MeV for the excited nucleus $^{96}$Mo. 

First, the present results of the $\alpha$-particle OMP \cite{va14} are in obvious agreement with the average trend of the data \cite{wr03} just above the discrete--resonance energy range \cite{exfor}. 
The extent of  this agreement could be pointed out by the factor of $\sim$5 between various OMP predictions formerly considered in this respect (Fig. 3 of Ref. \cite{wr03}). 
But more important is that also close to these data were the results provided by the first version \cite{ma03} of the OMP  \cite{va14}, i.e. based only on the $\alpha$-particle elastic--scattered analysis (Fig. 1 of Ref. \cite{ma06}). 
It is also worth noting that somewhat similar agreement shown in the latter figure by the quite different $\alpha$-particle  OMP \cite{va94}, which is only related only to neutron--induced $\alpha$-emission from $A$$\sim$60 nuclei. 
This fact may suggest that the simple OMP \cite{va94} could include, beyond the CN contribution, the additional ones that only now receive a full consideration. 

Second, almost for a complete picture, an eventual pickup DR contribution has been obtained by interpolating between the similar and comparable results for the target nuclei $^{92,98}$Mo using the SFs discussed in Sec.~\ref{DR}. 
As expected, DR cross--sections of less than 1 mb have a minor addition to the PE+CN sum, which is close to the early data around 14 MeV in the limit of twice the data standard deviation ($\sigma$), in Fig.~\ref{Fig:Mo92580na}(b). 
At the same time, a much larger underestimation of the more recent data around the incident energy of 5 MeV  \cite{gz10} is obvious. 
 
So, the third and most important is the suitable account of this extra-yield by the addition of a Gaussian distribution corresponding to the GQR energy of 14.2 MeV for excited nucleus $^{96}$Mo. 
A peak cross--section of 0.6 mb and a width of 2.35 MeV are given by the fit of the measured data \cite{gz10}. 
It is worth noting that there are no NLD and PE effects at these energies, the subsequent uncertainty bands becoming visible only at higher excitation. 
The DR contribution becomes also comparable with this {\it like}--GQR component only above an incident energy of 7 MeV. 
Therefore, it may be concluded that the data of Zhang {\it et al.} \cite{gz10} support the assumption of {\it like}--GQR $\alpha$-particle decay. 

{\it $^{98}$Mo$(n,\alpha)^{95}$Zr excitation function} [Fig.~\ref{Fig:Mo92580na}(c)] analysis reveals the pickup DR role within the completion of the CN+PE modeling.   
Thus, while CN+PE uncertainly bands due to the NLD and PE effects are visible from incident energies of 9--10 MeV, the DR contribution of SFs quoted in Sec.~\ref{DR} is significantly larger up to $\sim$13 MeV,  
Then, the uncertainty related to PE effects becomes more important above 14 MeV while the same happens for the NLD uncertainty band starting at 17 MeV. 
Nevertheless, due consideration of the DR has brought the agreement of the calculated cross--sections from the lower limit of the data error bars to their average values. 

On the other hand, the only data set \cite{mmr85} available below 10 MeV has shown an extra yield well above the DR+PE+CN cross--sections without either NLD or PE effects at these energies. 
These data could be described however by adding a Gaussian distribution corresponding to the GQR energy of 14.05 MeV for excited nucleus $^{99}$Mo, and the peak cross--section of 0.8 mb as well as the width of 2.35 MeV. % from the fit of the measured data \cite{mmr85}.  
The assumption of the {\it like}--GQR $\alpha$-particle decay appears to be supported once more. 
On the other hand, a fit of these data below 10 MeV, as seems to be the case for TENDL-2021 \cite{TENDL} in Fig.~\ref{Fig:Mo92580na}(c), is leading to a large overestimation of the measured data at higher energies.

{\it $^{100}$Mo$(n,\alpha)^{97}$Zr excitation function} [Fig.~\ref{Fig:Mo92580na}(d)] analysis is very similar to that of $^{98}$Mo, with the exception of no measured data below 10 MeV. 
The major drawback, however, of the lack of SFs needed for an accurate pickup DR cross-section estimation, has been overcome similarly to the case of $^{95}$Mo target nucleus. 
Thus, extrapolating the related components of $^{92,98}$Mo nuclei yielded results that are rather close to those for $^{98}$Mo. 
Finally, an unexpected agreement of the experimental data and the calculated CN+PE+DR sum has been found. 

To give a complete view of the possible {\it like}--GQR $\alpha$-emission for Mo isotopes, it is shown in Fig.~\ref{Fig:Mo92580na}(d) along with the outline of a Gaussian distribution at the GQR energy of 14.0 MeV for excited nucleus $^{101}$Mo. 
The width of 2.35 MeV and an assumed peak cross--section of 0.5 mb, close to that of $^{98}$Mo, have provided a possible shape to be confirmed or not by further measurements. 
Nevertheless, its proof could be favored by the isotope effect \cite{nim77} of significantly lower $(n,\alpha)$ cross--sections for heavier isotopes, similarly to Ni isotope chain \cite{va22}. 

\section{Conclusions}  \label{Conc}

A recent assessment of a previous $\alpha$-particle OMP \cite{va14} also for nucleon--induced $\alpha$-emission on $A$$\sim$60 nuclei, including pickup DR and eventual GQR $\alpha$-emission \cite{va21,va22}, is completed for neutrons incident on Zr, Nb, and Mo stable isotopes. 
Consistent sets of input parameters, determined through analysis of independent data, is involved %at variance with the use of either empirical rescaling factors of the $\gamma$ and neutron widths or combinations of all computer--code options for main input parameters.
with no further empirical rescaling factors of the $\gamma$ and nucleon widths which however are mandatory within large--scale nuclear--data evaluations. 
Nevertheless, there is an obvious correlation between the accuracy of the independent data, the input parameters determined by their fit, and final uncertainties of the calculated reaction cross sections. 
Moreover, additional validation of this potential is also supported by recently measured cross--sections of $(\alpha,\gamma)$ reactions on $^{90,91,92}$Zr as well as $(\alpha,n)$ on $^{96}$Zr and $^{100}$Mo nuclei. 

On the other hand, the pickup contributions to $(n,\alpha)$ reactions have been determined within the DWBA method using the code FRESCO \cite{FRESCO}. 
The one--step reaction has also been considered through the pickup of $^3$He cluster while the "spectator model" \cite{smits76,smits79} was involved for the two transferred protons in $(n,\alpha)$ reaction. 
However, the lack of measured $\alpha$-particle angular distribution for the $(n,\alpha)$ reactions within this work made possible only straightforward DWBA calculations of related pickup cross--sections. 
Thus, the spectroscopic factors of Glendenning \cite{glendenning} have been used for the spectator proton pair \cite{eg86,eg88,smits79}, in addition to SF for the picked neutron that becomes thus responsible for the angular--momentum transfer. 
The latter have been obtained through analysis of $\alpha$-particle angular distributions of one--nucleon pickup reactions $(^3He,\alpha)$ or $(t,\alpha)$ toward the residual nucleus of interest. 

Nonetheless, a suitable account by actual parameter set of all data for competing nucleon-emission by neutrons on Zr, Nb, and Mo isotopes has been a prerequisite for a consistent discussion of the related $\alpha$-emission. 
In this respect, the previous analyses for Zr \cite{va17} and Mo \cite{pr05} are completed by a similar work for $^{93}$Nb and newer measured data for the even--even semi--magic nucleus $^{92}$Mo.  
Then, an appropriate description of the  $\alpha$-particle angle--integrated energy distributions induced by neutrons on $^{93}$Nb, $^{90}$Zr, and $^{92}$Mo  has been concerned for validation of $\alpha$-particle PE and DR account. 

Finally, an increase of the $\alpha$-emission beyond the CN+PE predictions has been obtained through consideration of additional  pickup DR and {\it like}--GQR decay of excited nuclei by neutrons on Zr, Nb, and Mo stable isotopes. 
Thus it becomes possible a description of both the absorption and emission of $\alpha$ particles by the same potential \cite{va14}, in support of also its use for large--scale nuclear--data evaluations as TALYS corresponding default option. 
This conclusion had already been reachable within the previous analysis for Zr \cite{va17} and Mo \cite{ma06} before DR and {\it like}--GQR decay consideration, due to outstanding cross--section measurements for $^{91}$Zr$(n,n\alpha)^{87}$Sr$^m$ \cite{vs10} and $^{95}$Mo$(n,\alpha)^{92}$Zr \cite{wr03,gz10} reactions. 
At the same time it is suggested that the simple OMP \cite{va94} could include, beyond the CN contribution, the additional ones that only now receive full consideration. 
Nonetheless, further measurements at incident energies corresponding to GQR energies of excited nuclei, as well as heavier isotopes of elements, may shed light on the eventual {\it like}--GQR $\alpha$-emission. 

\section*{Acknowledgments}
This work has been partly supported by The Executive Unit for the Financing of Higher Education, Research, Development and Innovation (UEFISCDI) (Project No. PN-III-ID-PCE-2021-0642) and carried out within the framework of the EUROfusion Consortium, funded by the European Union via the Euratom Research and Training Programme (Grant Agreement No 101052200 — EUROfusion). Views and opinions expressed are however those of the author(s) only and do not necessarily reflect those of the European Union or the European Commission. Neither the European Union nor the European Commission can be held responsible for them.

\bibliography{mybibfileA2022}

\end{document}